\documentclass[times,twoside]{zHenriquesLab-StyleBioRxiv}
\usepackage{blindtext}
\usepackage{float}
\usepackage{xcolor}

\leadauthor{Becker} 

\begin{document}

\pagestyle{fancy}
\fancyhf{}
\fancyfoot[RE]{\footnotesize{Becker \textit{et al.} \hspace{0.1cm}| \hspace{0.1cm} Better than a lens - Increasing the signal-to-noise ratio through pupil splitting}}
\fancyfoot[LE]{\footnotesize{\thepage}}
\fancyfoot[LO]{\footnotesize{Becker \textit{et al.} \hspace{0.1cm}| \hspace{0.1cm} Better than a lens - Increasing the signal-to-noise ratio through pupil splitting}}
\fancyfoot[RO]{\footnotesize{\thepage}}

\title{\Large{Better than a lens} \\ \large{Increasing the signal-to-noise ratio through pupil splitting}}
\shorttitle{Better than a lens - Increasing the signal-to-noise ratio through pupil splitting}

\author[1,$\star$]{Jan Becker}
\author[2]{Takahiro Deguchi}
\author[1]{Alexander Jügler}
\author[1]{Ronny Förster}
\author[1]{\\Uwe Hübner}
\author[2]{Jonas Ries}
\author[1,3,4,$\dagger$]{Rainer Heintzmann}

\affil[1]{Leibniz Institute of Photonic Technology, Jena, Germany}
\affil[2]{Cell Biology and Biophysics Unit, European Molecular Biology Laboratory, Heidelberg, Germany}
\affil[3]{Institute of Physical Chemistry, Friedrich Schiller University, Jena, Germany}
\affil[4]{Abbe Center of Photonics, Friedrich Schiller University, Jena, Germany}
\affil[$\star$]{\emph{current affiliations:} Kavli Institute for NanoScience Discovery, Oxford, United Kingdom}
\affil[ ]{Physical and Theoretical Chemistry Laboratory, University of Oxford, United Kingdom}
\affil[$\dagger$]{\emph{corresponding author email:} \textrm{heintzmann@gmail.com}}

\newcommand{\amp}{\textit{amp}}
\newcommand{\NA}{\textit{NA}}
\newcommand{\RW}{\textit{RW}}
\newcommand{\WF}{\textit{WF}}
\newcommand{\SP}{\textit{SP}}
\newcommand{\Disk}{\textit{Disk}}
\newcommand{\Ring}{\textit{Ring}}
\newcommand{\SNR}{\textit{SNR}}
\newcommand{\eff}{\textit{ef}}
\newcommand{\NCC}{\textit{NCC}}
\newcommand{\IF}{\textit{IF}}
\newcommand{\FWHM}{\textit{FWHM}}
\newcommand{\BFP}{\textit{BFP}}
\newcommand{\CoM}{\textit{CoM}}
\newcommand{\Max}{\textit{max}}
\newcommand{\Min}{\textit{min}}
\newcommand{\Mod}{\textit{mod.}}
\newcommand{\Var}{\textit{Var}}
\newcommand{\Dim}{\textit{dim}}
\newcommand{\textcirc}[1]{\raisebox{.5pt}{\textcircled{\raisebox{-.9pt} {#1}}}}

\maketitle

\begin{abstract}
Lenses are designed to fulfill Fermat's principle such that all light interferes constructively in its focus, guaranteeing its maximum concentration. It can be shown that imaging via an unmodified full pupil yields the maximum transfer strength for all spatial frequencies transferable by the system. Seemingly also the signal-to-noise ratio (SNR) is optimal. The achievable SNR at a given photon budget is critical especially if that budget is strictly limited as in the case of fluorescence microscopy.  In this work we propose a general method which achieves a better SNR for high spatial frequency information of an optical imaging system, without the need to capture more photons. This is achieved by splitting the pupil of an incoherent imaging system such that two (or more) sub-images are simultaneously acquired and computationally recombined. We compare the theoretical performance of split pupil imaging to the non-split scenario and implement the splitting using a tilted elliptical mirror placed at the back-focal-plane (BFP) of a fluorescence widefield microscope.
\end {abstract}



\section{Introduction}
\label{sec:Introduction}

\begin{figure}[htb]
	\centering
	\includegraphics[width=0.95\linewidth]{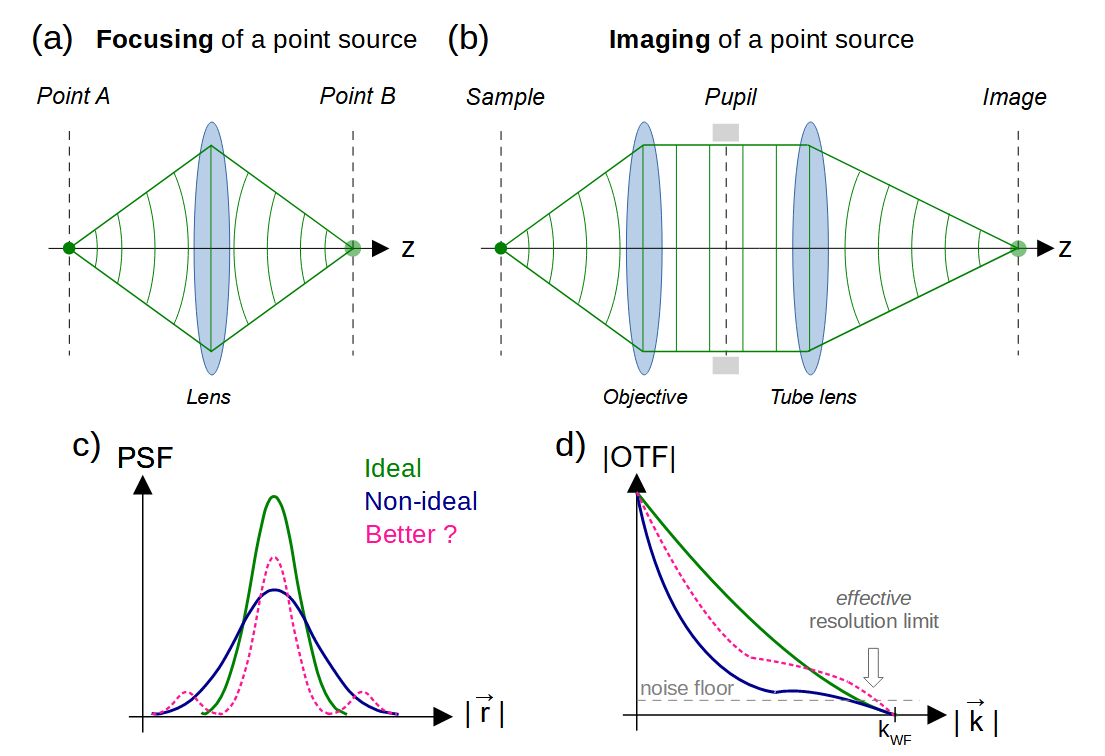}
	\caption{\textcolor{blue}{(a)} Fermat's principle predicts the shape of an ideal lens, which manipulates light such, that the emission at point \emph{A} is focused at point \emph{B}, where all rays mutually interfere constructively yielding maximum irradiance (supplement \ref{sec:Fermat's principle}). \textcolor{blue}{(b)} A $4f$-imaging system is characterized by the pupil function $\mathcal{\tilde{P}}$. \textcolor{blue}{(c)} The peak of the respective PSF $h$ is maximized for a purely transmissive $\mathcal{\tilde{P}}$, as any change in transmission or phase will inevitably lead to a peak-reduction (blue) of $h$ (supplement \ref{sec:Aberrated_Pupil}). \textcolor{blue}{(d)} A modification of $\mathcal{\tilde{P}}$ also reduces the absolute value of the OTF $\tilde{h}$ in Fourier space, indicating that it is not possible to go beyond the green curve by introducing an amplitude or phase changes to the detection pupil. However, when shot noise is considered we show that an improvement in terms of imaging high spatial frequency patterns can be achieved via pupil splitting \& computational recombination (depicted by the magenta curve).} 
	\label{fig:Fig1}
\end{figure}

Lenses have been around for many hundred years and well proven their qualities in forming images of
objects ranging from galaxies to single molecules. But are they always the best choice for the imaging task at hand? A lens is designed according to \emph{Fermat's principle} \cite{Mahoney_Fermat}: light emitted at a point A is focused by the lens into a point B. All optical paths from A to B have the same lengths causing constructive interference at B (supplement \ref{sec:Fermat's principle}), which yields a maximum irradiance of light in B \cite{Feynman_Lectures1} (Fig. \ref{fig:Fig1} \textcolor{blue}{(a)}). In imaging using a $4f$ setup (Fig. \ref{fig:Fig1} \textcolor{blue}{(b)}) the system performance can be summarized by the point-spread-function $h$ (\ref{fig:Fig1} \textcolor{blue}{(c)}; also see supplement \ref{sec:IncoherentImageFormation}) and its Fourier transform $\tilde{h}$, the optical transfer function (OTF; Fig. \ref{fig:Fig1} \textcolor{blue}{(d)}), describing how well a particular spatial frequency of the incoherent object is transferred by the imaging system. Whenever the transmission or phase distribution of the pupil $\mathcal{\tilde{P}}$ is changed, this will decrease the peak value of $h$ \cite{Goodman_Fourier}. Interestingly an even more general effect is that the OTF $\tilde{h}$ will decrease at some spatial frequencies without the potential for any increase as long as the pupil modifications cannot amplify the light (Fig. \ref{fig:Fig1} \textcolor{blue}{(d)}). This yields a worse imaging performance whenever $\mathcal{\tilde{P}}$ is manipulated (supplement \ref{sec:Aberrated_Pupil}).
Let us now consider the influence of photon noise on the imaging performance. Light consists of particles (photons) which are measured in counts, leading to \emph{shot} noise \cite{Mertz_Book}. Neglecting all other noise sources (e.g. readout and dark noise), the measurable image pixel intensities are \emph{Poisson} distributed, with the noise variance being proportional to the expected signal \cite{Lucke_PoissonNoise}. In Fourier space shot noise appears as a constant noise floor (supplement \ref{sec:Photon-limited noise}), which is proportional to the total number of expected photons. The frequencies at which the OTF touches the noise floor yields an effective resolution limit for a given photon budget. A reduction in photon flux decreases noise, albeit also lowering the signal level. A given photon budget limits the achievable signal-to-noise ratio ($\SNR \propto \sqrt{N_{\text{photons}}}$ ). Which can only be overcome when more photons are being detected (supplement \ref{sec:SNR_Real_Fourier}), often causing other detrimental effects such as reduced temporal resolution, increased photobleaching or phototoxicity \cite{Stelzer_SNR}. In this work we demonstrate that the SNR can be improved, without the need to capture more photons. This is achieved by splitting the pupil of a detection objective, image the sample via each sub-pupil and computationally recombine the acquired sub-images to achieve such SNR improvement. The splitting is advantageous as the noise in both sub-images will be reduced due to the decreased photon number, while some of the signal information, especially the high-resolution information, is maintained. The computational reconstruction is done using weighted averaging in Fourier space \cite{Wicker_PhDThesis,Becker_Better,Heintzmann_Subtraction} or via multiview deconvolution \cite{York_MultiviewDeconvolution,Heintzmann_AxialTomography}, both extracting more sample information from the limited photon budget. 

\section{Results}
\label{sec:Results}

\subsection*{Pupil splitting \& recombination}
\label{sec:Pupil_splitting}

\begin{figure}[htb]
	\centering
	\includegraphics[width=\linewidth]{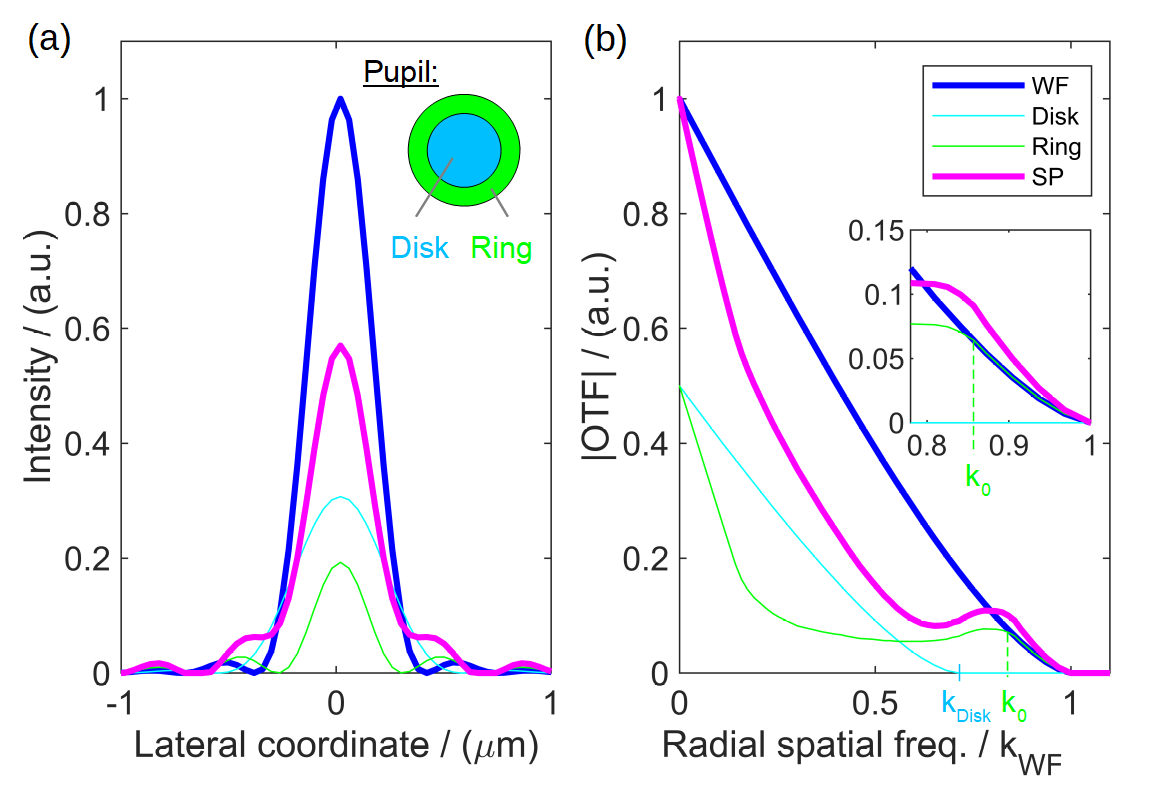}
	\caption{\textcolor{blue}{(a)} Individual (simulated) PSFs corresponding to imaging with a WF (blue), inner disk (cyan) or outer ring (green) pupil. In case of pupil \emph{splitting \& recombination} (SP), the (effective) weighted averaged result is shown in magenta color. \textcolor{blue}{(b)} Corresponding sub-OTFs of aforementioned pupil settings. For $|\vec{k}| \geq k_0$ we observe that $|\tilde{h}_\Ring| = |\tilde{h}_\WF|$. Meaning that less photons are needed (only $\eta^2 = 50 \%$) to create the same signal strength for those particular frequencies (see supplement \ref{sec:OTF_autocorrelation}). Note that the recombined result (magenta) is noise-normalized and goes beyond the conventional WF curve at frequencies for $|\vec{k}| \gtrsim k_0$. Details on how the individual curves have been calculated are given in Methods \ref{sec:Computing_OTFs}.}
	\label{fig:Fig2}
\end{figure}

In Fourier space the performance of an incoherent imaging system is given by the OTF $\tilde{h}$. Which is, due to the incoherent PSF being $h=|h_{\amp}|^2$ (with $h_{\amp}$ the inverse Fourier transform of the pupil function $\mathcal{\tilde{P}}$), computed as the autocorrelation $\mathcal{A}\{\mathcal{\tilde{P}}\}$ \cite{Goodman_Fourier}. Given in the following in a scalar form:
\begin{equation}
	\tilde{h}(\vec{k}) = \mathcal{A} \big \{ \mathcal{\tilde{P}}(\vec{k}) \big \} = \int_{-\infty}^{+\infty} d\vec{k^\prime} \hspace{4pt} \mathcal{\tilde{P}}(\vec{k^\prime}) \cdot  \mathcal{\tilde{P}}^*(\vec{k^\prime}-\vec{k})
	\label{eq:Autocorrelation}
\end{equation}
with $^*$ representing the complex conjugate and $\vec{k}$ spatial frequency. Note that quantities in Fourier space are denoted with a tilde, e.g. $\tilde{x}$. The OTF of an imaging system that employs a full pupil (widefield imaging) is denoted with $\tilde{h}_{\WF}$. We radially split the widefield pupil $\mathcal{\tilde{P}}_{\WF}$ (with a cutoff frequency $k_{\WF}$ \cite{Abbe_Beitraege}), resulting in two sub-pupils with the shape of a disk $\mathcal{\tilde{P}}_{\Disk}$ (with cutoff $k_{\Disk}$) and a ring $\mathcal{\tilde{P}}_{\Ring}$. In terms of signal transfer it can be shown that spatial frequencies beyond $k_0 = 0.5 \cdot (k_{\WF} + k_{\Disk})$ are maintained when imaged through the ring pupil, instead of $\mathcal{\tilde{P}}_\WF$ (see supplement \ref{sec:OTF_autocorrelation}):
\begin{equation}
	\tilde{h}_{\WF}\left(|\vec{k}| \geq k_0\right) = \tilde{h}_{\Ring} \left(|\vec{k}| \geq k_0\right)
\end{equation}
This effect is depicted in Fig. \ref{fig:Fig2}, which shows the individual PSFs \textcolor{blue}{(a)}  and OTFs \textcolor{blue}{(b)} for the widefield (blue), disk (cyan) and ring (green) pupil. The radial split is parameterized by $\eta = k_{\Disk} / k_{\WF} = R_{\Disk} / R_{\WF}$ with $R_{\Disk,\WF}$ being the radii of the actual pupil stops in the optical system.
In the following, the split was set to $\eta = \sqrt{0.5}$, which yields two sub-pupils, $\mathcal{\tilde{P}}_\Disk$ and $\mathcal{\tilde{P}}_\Ring$, with equal areas.

Hence the corresponding two sub-OTFs (Fig. \ref{fig:Fig2}, cyan and green) exhibit a value of $0.5$ at $\vec{k} = 0$ (half the photon numbers in each pupil compared to the WF case). The green curve shows that it is possible to transfer high spatial frequency information ($|\vec{k}| \geq k_0$) with the same signal strength as in the non-split pupil case (WF), while only a fraction (here $\eta^2 = 0.5$) of photons are required. In terms of SNR (in Fourier space; see supplement \ref{sec:SNR_Real_Fourier}), this indicates an improvement of $1 / \sqrt{1 - \eta^2}$ (here $41 \%$) at those particular frequencies, while for $|\vec{k}| < k_0$ the SNR is reduced. The latter can be partly overcome when the image data corresponding to the ring pupil (green) is computationally combined with the one related to the disk pupil (cyan), as $\mathcal{\tilde{P}}_{\Disk}$ carries more information about the low to medium spatial  frequencies.\\
One approach to recombine both sub-images is \emph{weighted averaging} in Fourier space, which fuses the two images according to their spectral representation $\tilde{I}_{1,2}$:
\begin{equation}
	\tilde{I}_{wa }(\vec{k}) = \tilde{w}_1(\vec{k}) \cdot \tilde{I}_1(\vec{k}) + \tilde{w}_2(\vec{k}) \cdot \tilde{I}_2(\vec{k})
\end{equation}
with the subscripts indicating the information corresponding to the disk \& ring pupil respectively. The individual (noise-normalized) weights $\tilde{w}_l$ are given according to \cite{Wicker_PhDThesis,Becker_Better,Heintzmann_Subtraction}:
\begin{equation}
	\tilde{w}_l(\vec{k}) = \dfrac{\tilde{h}_l(\vec{k}) / \tilde{\sigma}_l^2}{\sqrt{\tilde{h}_1^2(\vec{k}) / \tilde{\sigma}_1^2 + \tilde{h}_2^2(\vec{k}) / \tilde{\sigma}_2^2}}
\end{equation}
with $\tilde{\sigma}_{1,2}^2$ being the Fourier noise variance of the respective sub-images. The weights are chosen such that for each frequency $\vec{k}$, information of both images are added in a SNR optimal way (supplement \ref{sec:WeightedAvg}). As this computation is linear, the overall process of imaging and computationally recombining can be summarized as an effective OTF $\tilde{h}_{wa }$:
\begin{equation}
	\tilde{h}_{wa} (\vec{k}) = \sqrt{\dfrac{\tilde{h}_1^2(\vec{k})}{\tilde{\sigma}_1^2} + \dfrac{\tilde{h}_2^2(\vec{k})}{\tilde{\sigma}_2^2}}
\end{equation}
The corresponding OTF curve (and the respective PSF) is shown in Fig. \ref{fig:Fig2} in magenta color, indicating an SNR improvement at high spatial frequencies. This creates a slightly enhanced effective resolution (where $\tilde{h}_{wa }$ reaches the noise floor, see Fig. \ref{fig:Fig1}), while the transfer strength at mid-frequencies is reduced. Note that the latter is a drawback we can accept, as long as the corresponding image information stays above the noise floor. The missing transfer strength in the pupil splitting approach can still be corrected by using post-processing, while this is not advisable for high spatial frequency information which has fallen below the noise floor.\\
Another way to perform image fusion is via maximum likelihood based iterative \emph{multiview deconvolution}, e.g. estimating the underlying sample $S$ from the two sub-images $I_{1,2}$ \cite{Heintzmann_AxialTomography}. We use an iterative Richardson-Lucy algorithm (RL) \cite{York_MultiviewDeconvolution}, where in each iteration the current estimate $\hat{S}_j$ is multiplied with a correction factor $C_j$ \cite{Richardson_Deconvolution,Lucy_Deconvolution}: 
\begin{equation}
	\hat{S}_{j+1}(\vec{r}) = \hat{S}_j(\vec{r}) \cdot C_j(\vec{r})
\end{equation}
with $\vec{r}$ being real space coordinates and $j$ the current iteration number. The conventional RL correction factor is modified to yield a \emph{weighted} multiview RL version, given by:
\begin{equation}
	C_j(\vec{r}) = \mathcal{F}^{-1}  \Bigg\{ \sum_{l=1}^2 \tilde{w}_{l}(\vec{k})\cdot \mathcal{F}\bigg\{  \dfrac{I_l(\vec{r})}{\hat{S}_j(\vec{r}) \otimes h_l(\vec{r})} \otimes h_l(-\vec{r}) \bigg\} \Bigg\}
\end{equation}
with $\mathcal{F}$ being the Fourier transform operator. See supplement \ref{sec:Multiview_Deconvolution} for a more detailed description of this multiview algorithm. Throughout this work we have used the accelerated version of the RL algorithm described in Biggs et al. \cite{Biggs_RLAcceleration}.

\subsection*{Numerical results}
\label{sec:Numerical_results}

To theoretically investigate the dependency of the possible enhancement in terms of SNR, achievable with linear processing (weighted averaging), the improvement factor $\IF$ is introduced. Which is the relative difference between the weighted averaged OTF $\tilde{h}_{wa}$ (split \& recombined pupil) and the conventional widefield version $\tilde{h}_{\WF}$ (non-split pupil):
\begin{equation}
	\IF(\vec{k}) = \dfrac{|\tilde{h}_{wa} (\vec{k})|}{|\tilde{h}_{\WF}(\vec{k})|} - 1
\end{equation}
An improvement factor of $\IF(\vec{k}) = 100\%$ relates to a doubling of the signal strength at the frequency $\vec{k}$. The distribution of $\IF$ in Fourier space is shown in Fig. \ref{fig:Fig3} \textcolor{blue}{(a)} and in supplement \ref{sec:ImprovementFactor} for different splitting radii (related to $\eta$).

\begin{figure}[htb]
	\centering
	\includegraphics[width=\linewidth]{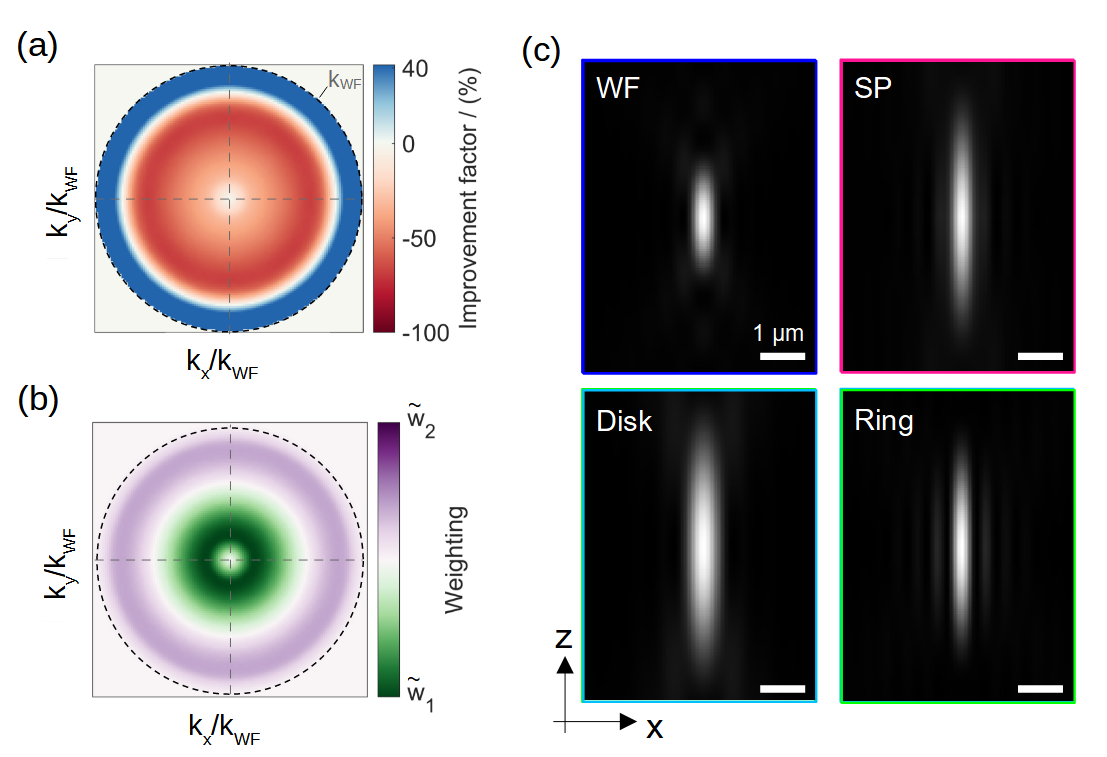}
	\caption{\textcolor{blue}{(a)} Improvement factor $\IF$ (in \emph{percent}) at different spatial frequencies $\vec{k}$ for $\eta = \sqrt{0.5}$. The maximum and minimum achievable improvement are $\IF_{\Max} \approx 42 \%$ and $\IF_{\Min} \approx -68 \%$ (see Tab. \ref{tab:Tab1}), respectively. The change of $\IF$ with respect to $\eta$ is depicted in Fig. \ref{fig:FigA2b}. \textcolor{blue}{(b)} Visualization of $|\tilde{w}_1| - |\tilde{w}_2|$ used in the weighted averaging recombination approach. The dashed circle indicates the $|\vec{k}|/k_{\WF} = 1.0$. Note that high spatial frequency information ($|\vec{k}| > k_\Disk$) is only taken from the sub-image corresponding to $\mathcal{\tilde{P}}_\Ring$.  \textcolor{blue}{(c)} $XZ$-view on the different (normalized) PSFs indicating an extended depth-of-field (EDoF, see supplement \ref{sec:EDoF_effect}) of the recombined result (top right), compared to the WF case.}
	\label{fig:Fig3}
\end{figure}

As already suggested in Fig. \ref{fig:Fig2} \textcolor{blue}{(a)}, a positive improvement is only achievable for high spatial frequencies ($|\vec{k}| \geq k_0$), while medium frequencies experience a "negative improvement". The corresponding maximum or minimum improvement factor $\IF_{\Max,\Min}$ and the fraction of Fourier space which obtains a positive improvement $\IF>0$, are given in Table \ref{tab:Tab1} for a range of radial splitting parameters $\eta$:

\begin{table}[H]
	\centering
	\small{
		\begin{tabular}{|l|ccccc|}
			\hline
			$\eta$
			& 0.25 & 0.50 & 0.75 & 0.95 & $\sqrt{0.5}$ \\
			\hline
			$\IF_{\Max}$                                                                   
			& 3.20 & 15.26 & 51.10 & 201.17 & 41.65 \\
			$\IF_{\Min}$                                                                      
			& -21.05 & -57.32 & -68.95 & -68.72 & -68.48 \\
			$\IF>0$                                                                               
			& 63.16 & 50.08 & 31.27 & 8.18 & 35.13 \\
			\hline    
	\end{tabular}}
	\caption{Maximum and minimum improvement factor $\IF_{\Max,\Min}$ and area fraction in Fourier space with a positive improvement ($\IF>0$) in dependence of on the pupil splitting parameter $\eta = R_\Disk / R_\WF$. All values are given in \emph{percent}. }
	\label{tab:Tab1}
\end{table}

The maximum achievable improvement $\IF_{\Max}$ scales with $\eta$, meaning that a strong SNR enhancement requires the ring pupil to be narrow. However, this also restricts the region of improvement $\IF>0$ to only very high spatial frequencies (close to $k_{\WF}$). Hence, a tradeoff between enhancement strength and number of improvable spatial frequencies needs to be found. In this paper we have set  $\eta = \sqrt{0.5}$, yielding in an maximum improvement of $\IF_{\Max} \approx 42 \%$, while an area fraction of $\approx 35\%$ within the detection band limit are being enhanced. The corresponding weights are depicted in Fig. \ref{fig:Fig3} \textcolor{blue}{(b)} as the difference $|\tilde{w}_1| - |\tilde{w}_2|$. High spatial frequency information ($|\vec{k}|>k_\Disk$) is taken solely from the sub-image corresponding to the ring pupil. Note that the signal strength at the mid-frequencies is not lost and can still be recovered with appropriate image reconstruction methods, as $\IF_{\Min} >-100 \%$ for all $\vec{k}$. This is exemplified in supplement \ref{sec:Multiview_Deconvolution}, which shows reconstruction results after performing the aforementioned multiview deconvolution approach to recombine simulated split pupil image data of a spokes target. The multiview RL-method achieves an enhanced reconstruction at high spatial frequencies, while the reduced transfer strength at the mid-frequencies is recovered.\\
Another aspect to keep in mind is that the radial pupil split will also alter the imaging performance along the axial direction. This is because the corresponding sub-PSFs are that of a WF system with a reduced numerical aperture ($\NA_\Disk = \eta \cdot \NA_\WF$) and a ring pupil, which both show an elongated axial extend (compared to $h_{\WF}$). This leads to an extended depth-of-field (EDoF) effect for split pupil imaging as shown in Fig. \ref{fig:Fig3} \textcolor{blue}{(c)} for $\eta=\sqrt{0.5}$ (extending the axial range of $h_{wa }$ by a factor of $\approx 2$). A more detailed explanation is given in supplement \ref{sec:EDoF_effect}.

\subsection*{Experimental results}
\label{sec:Experimental_results}

The pupil splitting was experimentally realized in a widefield imaging setup, by placing an elliptical mirror on a glass substrate in the BFP of the detection objective (see Fig. \ref{fig:Fig2} in Methods \ref{sec:Custom_EllipticalMirror}). Note that we removed the mirror whenever we recorded widefield data, so that a fair comparison with respect to conventional imaging (i.e. full pupil) was possible. The BFP was accessed by including an additional lens behind the nominal image plane of the optical system, where the mirror reflects light that corresponds to $\mathcal{\tilde{P}}_\Disk$, while the transmitted light refers to $\mathcal{\tilde{P}}_\Ring$. Both channels form an image, side-by-side, on the same camera (additional lens in each arm), simplifying simultaneous image acquisition. The diameter of the mirror has been chosen to yield the aforementioned "equal-area" splitting, corresponding to $\eta = 0.68 \approx \sqrt{0.5}$ for the used water immersion objective ($\NA = 1.2$).\\
To verify the capability of high spatial frequency signal enhancement, we imaged a fluorescent resolution pattern from an Argolight calibration target \cite{Royon_Argolight} (Argo-SIM, Argolight SA, Pessac, France). It consists of line pairs separated by a distance (spacing $\Delta s$) ranging from $0$ to $390$ nm, enabling us to see the change in modulation for different relative spatial frequencies $|\vec{k}|/k_{\WF}$. A detailed description on the data acquisition and evaluation are given in Methods \ref{sec:Experimental_setup}. To quantify the achieved improvement, each measurement consists of a time-series (100 images), enabling us to compute the expectation value $\mu$  (= mean) and standard deviation $\sigma$ over the recorded as well as the processed data. The expectancy of the split sub-images, as well as the widefield and recombined image results are depicted in supplement \ref{sec:Image_results_ResolutionTarget}, before and after deconvolution. Figure \ref{fig:Fig3} shows a line profile for the deconvolved widefield (blue; 24 iterations) and split \& recombined image results (magenta; 18 iterations - for difference in iteration number see supplement \ref{sec:Multiview_Deconvolution}).

\begin{figure}[htb]
	\centering
	\includegraphics[width=\linewidth]{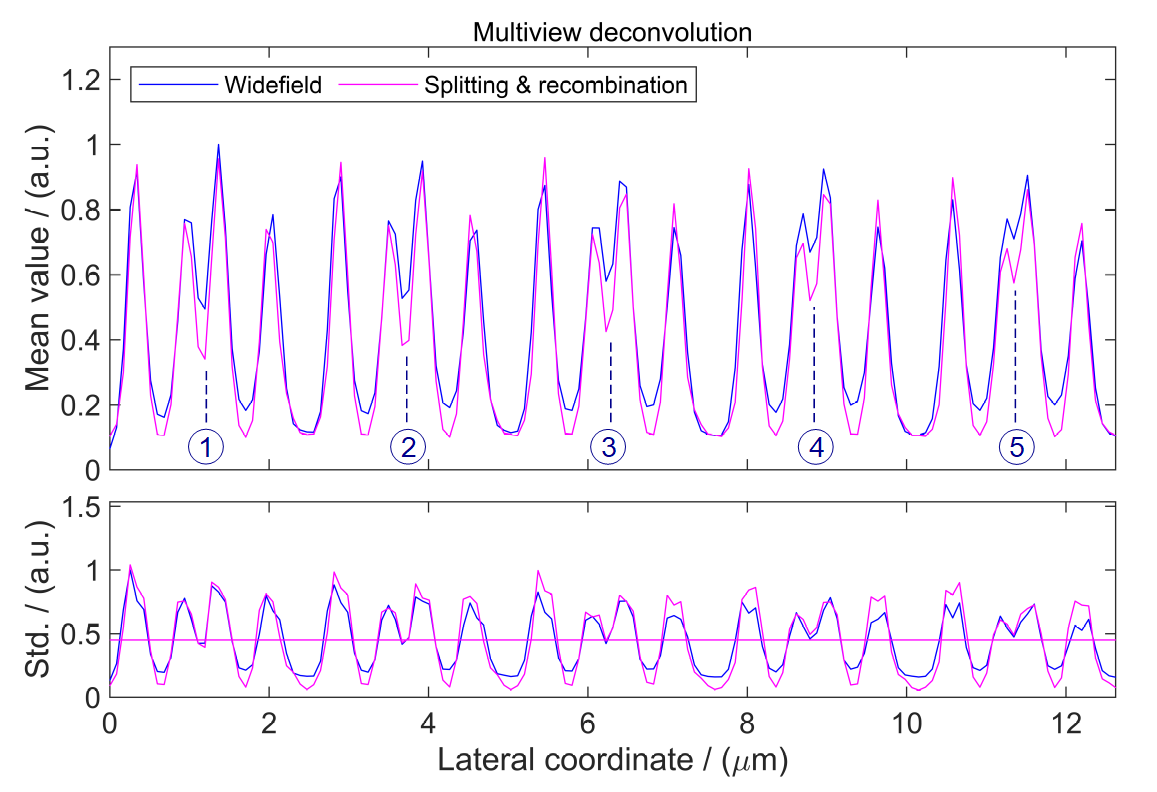}
	\caption{Line profile through the imaged Argolight \cite{Royon_Argolight} resolution pattern, showing the mean value (top) and standard deviation (bottom) after processing of a time-series of 100 recorded images (circled numbers correspond to line pairs evaluated in Tab. \ref{tab:Tab2}). Both imaging modalities (widefield: blue, without splitting mirror; splitting \& recombination: magenta, with splitting mirror) are noise-normalized so that their modulation strength (or visibility $\mathcal{V}$) is a direct indicator of the attainable SNR enhancement. The shown results are deconvolved using the aforementioned weighted RL algorithm (Iterations: WF= 24; SP= 18).}
	\label{fig:Fig4}
\end{figure}

Note that the recombined result is noise-normalized with respect to the WF case, which is indicated by the same average standard deviation (horizontal line; in Fig. \ref{fig:Fig4} bottom). Due to the band-limited nature of microscopic imaging, the resolvable modulation of the line pairs gets smaller for decreasing line separation $\Delta s$. The improved SNR of pupil splitting yields a stronger modulation for line patterns corresponding to $|\vec{k}|>k_0$, i. e. (\textcirc{3} - \textcirc{5}). This is verified in Table \ref{tab:Tab2}, by stating the visibility $\mathcal{V}$ as a direct indicator for the SNR enhancement (more on the computation of $\mathcal{V}$ in Methods \ref{sec:Calculating_visibility}). The corresponding visibility improvement factor $\IF_\mathcal{V} = \mathcal{V}_\SP / \mathcal{V}_\WF -1$ is stated for the weighted average (WA) and multiview deconvolution (MV) recombination.

\begin{table}[htb]
	\centering
	\small{
		\begin{tabular}{|c|l|>{\centering}m{0.85cm}>{\centering}m{0.85cm}>{\centering}m{0.65cm}>{\centering}m{0.65cm}>{\centering\arraybackslash}m{0.65cm}|}
			\hline
			&  Line pair  & \textcirc{1} & \textcirc{2} & \textcirc{3} & \textcirc{4} & \textcirc{5}\\ \hline 
			&  $\Delta s$ / (nm)  & 390 & 360 & 330 & 300 & 270\\ 
			& $|\vec{k}|/k_{\WF}$ & 0.56 & 0.60 & 0.66 & 0.72 & 0.80 \\
			\hline
			\multirow{3}{*}{\scriptsize{\emph{WA}}}  & $\mathcal{V}_{\textcolor{blue}{\WF}}$ & 0.060 & 0.042 & 0.027 & 0.000 &  0.000 \\
			& $\mathcal{V}_{\textcolor{magenta}{\SP}}$ & 0.042 & 0.037 & 0.031 & 0.019 & 0.014 \\
			& $\IF_\mathcal{V}$ / (\%) & \textit{-28.69} & \textit{-11.41} & \textit{14.36} &\textit{$\infty$} & \textit{$\infty$}  \\
			\hline
			\multirow{3}{*}{\scriptsize{\emph{MV}} } & $\mathcal{V}_{\textcolor{blue}{\WF}}$ & 0.283 & 0.238 & 0.168 & 0.123  & 0.083                                    \\
			&  $\mathcal{V}_{\textcolor{magenta}{\SP}}$ & 0.431 & 0.371 & 0.298 & 0.238 & 0.146 \\
			& $\IF_\mathcal{V}$ / (\%)  & \textit{32.40} & \textit{37.00} & \textit{53.28} & \textit{74.15} & \textit{52.21} \\
			\hline              
	\end{tabular}}
	\caption{The visibility $\mathcal{V}$ for five different line separations $\Delta s$ shown in Fig. \ref{fig:Fig4} (circled numbers, $k_\WF \approx 210$ nm). The recombination of the split data was either done by weighted averaging (WA) or multiview deconvolution (MV; widefield = 24, split pupil = 18 iterations). The visibility \emph{improvement} factor is given as $\IF_\mathcal{V} = \mathcal{V}_\SP / \mathcal{V}_\WF -1$. Note that in the non-deconvolved case the two line patters, corresponding to $\textcirc{4}$ and $\textcirc{5}$, could only be resolved using the split pupil approach (hence marked with $\IF_\mathcal{V} = \infty$). Details on the visibility computation are given in Methods \ref{sec:Calculating_visibility}.}
	\label{tab:Tab2}
\end{table}

In the case of the weighted average recombination (WA, see Fig. \ref{fig:FigA5a}), the first two line pairs (\textcirc{1} \& \textcirc{2}) show a negative improvement. Which is to be expected as the transfer strength of $\tilde{h}_{wa }$ is much reduced for those frequencies (magenta in Fig. \ref{fig:Fig1}). Nevertheless, when decreasing the line pair separation $\Delta s$ an improvement of $\approx 14  \%$ becomes observable.
\if
From the theory we would expect to have an equal performance at $k_0/k_\WF \approx 0.85$, which is not the case in the experiment. One reason for this could be the \emph{aplanatic factor}, whose influence is investigated in supplement \ref{sec:Aplanatic_factor}. 
\fi
For the highest spatial frequencies (\textcirc{4} \& \textcirc{5}) the split pupil approach was still able to resolve the line pair structure, while widefield imaging failed to do so (marked as $\IF_\mathcal{V} = \infty$). Recombining the data using multiview deconvolution (MV, see Fig. \ref{fig:Fig4}) generated a stronger improvement, and more importantly, an enhancement for all line pairs (compared to the deconvolved WF result). Showing that the reduced transfer strength at medium frequencies, of split pupil imaging, can be overcome using data post-processing. The enhancement is on the order of $\approx 30 - 70 \%$, indicating the inherit improvement of SNR in the split pupil approach. To experimentally evaluate the EDoF ability of pupil splitting, the "stair"-target (same cylindrical object at different depths) of the Argolight sample \cite{Royon_Argolight} has been imaged. Raw images and the deconvolved results are given in supplement \ref{sec:Image_results_StairsTarget}.

\section{Discussion}
\label{sec:Discussion}

We have shown that, contrary to common belief, it does make sense to manipulate the pupil of an incoherent imaging system to obtain \emph{better} imaging performance in terms of SNR. Instead of changing the amplitude or phase information, we remove the interference capability of light by splitting the pupil into two separate regions, yielding two sub-images. Interestingly the splitting does not change the transfer strength for high spatial frequencies, while only a fraction of the original photon budget is required. Subsequent computational recombination enables to extract more information from a finite number of photons, compared to imaging with a non-split pupil. Modifying the pupil of an imaging system is not novel when aiming for an improvement in imaging performance. Previous pupil modification methods, however, often reduced imperfection in non-ideal detection systems (e.g. reduce sidelobes by apodization \cite{McDonald_Apodization} or avoid aberrations by shrinking the pupil \cite{Barakat_Apodization_Coherent,Barakat_Apodization_Incoherent}). Instead we aim to improve the SNR beyond the seemingly unsurpassable limit dictated by the full pupil OTF \cite{Becker_Better}.\\
In many practical cases there may still be other options to improve the SNR, such as increasing the exposure time, getting a more efficient detector or enhancing the detection NA or the setup. Yet if such options are not available or have already been used, the split pupil imaging approach is still able to enhange the SNR. Increasing the $\NA$ is only possible until the maximum $\NA$ of commercially available objectives is reached, whereas pupil splitting generates an SNR improvement without the need to actually collect more photons. Similar arguments can be made for any other traditional method which improves SNR, such as increasing the exposure time of the detector or raising the illumination power on the sample. Furthermore, any such method comes with a drawback, e.g. increasing exposure time decreases temporal resolution.\\ 
Of course our proposed method also comes with a caveat: it requires double the field-of-view (FoV), if both sub-images are captured on the same sensor. Fortunately it is just as well possible to use two seperate cameras running synchronously. Issues such as synchronization and proper calibration are more practical and can certainly be overcome.\\ 
Throughout our work we have assumed that additional noise sources can be neglected. Modern cameras exhibit a readout noise level in the range of $1e^-$ RMS (root-mean-square) at a frame rate of 20 - 500 kHz \cite{VanVliet_CameraNoise}. In our measurements the frame rate was on the order of 1 kHz. Acquiring two instead of just one image yields a $\sqrt(2)$ larger read out noise at the worst weighted averaged frequency. However, even in the photon-limited regime, the generated number of photo-electrons usually is $\gg 1$. Fixed pattern noise and the detector offset have been removed by subtracting dark images, corresponding to zero photon flux at the detector. Therefore our assumption to neglect readout and dark noise in our analysis\\
The main advantage of our proposed method, is that it is possible to achieve an SNR enhancement without losing temporal resolution or requiring an increase in illumination, making it ideal for a wide range of applications. Especially in the field of metrology where an imaging task can often be defined very specific (e.g. the identification of certain manufacturing defects), in which case tailored pupil splits (e.g. by employing a spatial-light-modulator, as used in \cite{Becker_Better}) may lead to very large relative SNR improvements, allowing a significant reduction of exposure time. Another interesting application field could lie in astronomy where usually small objects (in terms of their angular extend in the sky), such as stars or planets should be resolved. These objects correspond to high spatial frequency information, which would especially benefit from split pupil imaging (which is independent of the detection wavelength). Together with adaptive optics it may even be possible to demonstrate an adaptive split pupil approach, where the split is adjusted depending on the structure of the object under study.

\section{Methods}
\label{sec:Methods}
\small{
	
\subsection{Computing OTFs from pupil functions}
\label{sec:Computing_OTFs}

Instead of calculating the individual OTFs directly from the pupil function using an autocorrelation operation (as suggested in eq. \ref{eq:Autocorrelation}), we first compute the PSF and Fourier transform ($\mathcal{F}$) the result, to obtain the OTF.
\begin{equation}
	\tilde{h}(\vec{k}) = \mathcal{F} \big\{ h(\vec{r}) \big\}
\end{equation}
A definition of $\mathcal{F}$ is given in supplement \ref{sec:IncoherentImageFormation}. We have chosen to start our calculation in real space, because this is the typical result of many accurate PSF models. Such models encompass effects such as focusing under high angles, the aplanatic factor and the polarization of light. For our simulation we used the traditional method of Richards \& Wolf (RW) \cite{RichardsWolf_VecPSF} with input parameters: numerical aperture ($\NA=1.2$), refractive index of immersion ($n=1.333$), wavelength ($\lambda = 520$ nm) and the  polarization state (unpolarized) by defining a pupil function in the BFP. The RW method (denoted by $f_{\RW}$) is able to calculate the PSF for a widefield system in terms of its electric field, the amplitude-spread-function $a$ (shown here, similar to sec. \ref{sec:Results}, in a scalar form).
\begin{eqnarray}
	a_{\WF}(\vec{r}) &=& f_{\RW}(\NA, \hspace{2pt}n,\hspace{2pt} \lambda, \hspace{2pt} \mathcal{\tilde{P}}_{\WF}) \\
	a_{\Disk}(\vec{r}) &=& f_{\RW}(\NA, \hspace{2pt} n,\hspace{2pt} \lambda, \hspace{2pt}\mathcal{\tilde{P}}_{\Disk}) 
\end{eqnarray}
The corresponding intensity PSFs are given according to:
\begin{eqnarray}
	h_{\WF}(\vec{r}) &=& \big| a_{\WF}(\vec{r}) \big|^2  \label{eq:PSF_WF}\\
	h_{\Disk}(\vec{r}) &=& \big| a_{\Disk}(\vec{r}) \big|^2 \label{eq:PSF_Disk}
\end{eqnarray}
To calculate the PSF corresponding to the ring pupil we use the fact that $\mathcal{\tilde{P}}_{\Ring}$ can be expressed in terms of $\mathcal{\tilde{P}}_{\WF}$ and $\mathcal{\tilde{P}}_{\Disk}$:
\begin{equation}
	\mathcal{\tilde{P}}_{\Ring}(\vec{k}) = \mathcal{\tilde{P}}_{\WF}(\vec{k}) - \mathcal{\tilde{P}}_{\Disk}(\vec{k})
\end{equation}
In general, the amplitude-spread-function $a$ is given as the inverse Fourier transform $\mathcal{F}^{-1}$ (defined in supplement \ref{sec:IncoherentImageFormation}) of the pupil:
\begin{equation}
	a(\vec{r}) = \mathcal{F}^{-1} \big\{ \mathcal{\tilde{P}}(\vec{k}) \big\}
\end{equation} 
Which means that $a_{\Ring}$ is given as the subtraction:
\begin{eqnarray}
	a_{\Ring}(\vec{r}) = a_{\WF}(\vec{r}) - a_{\Disk}(\vec{r})
\end{eqnarray}
Hence, the corresponding intensity point-spread-function can be directly computed from $a_{\WF}$ and $a_{\Disk}$ according to: 
\begin{eqnarray}
	h_{\Ring}(\vec{r}) = \big| a_{\WF}(\vec{r}) - a_{\Disk} (\vec{r}) \big|^2 \label{eq:PSF_Ring}
\end{eqnarray}
So that $f_{\RW}$ can be used to calculate $h_{\WF}$ and $h_{\Disk}$ directly through eq. \ref{eq:PSF_WF} - \ref{eq:PSF_Disk} and $h_{\Ring}$ indirectly via eq. \ref{eq:PSF_Ring}.

\subsection{Characterizing the custom elliptical mirror}
\label{sec:Custom_EllipticalMirror}

The pupil splitting is achieved using a mirror device placed in a plane conjugate to the BFP of a microscope objective. As can be seen from Fig. \ref{fig:Fig5} the mirror is placed under an angle ($\approx 10^\circ$) reflecting the light distribution corresponding to $\mathcal{\tilde{P}}_{\Disk}$ into one arm. Due to the tilt the shape of the mirror is elongated along one axis, so that the projection remains circular. As the shape and size of the mirror was very specific to our implementation of pupil splitting, we manufactured it ourselves with the help of the \emph{Competence Center of Micro- and Nanotechnologies} at the Leibniz Institute of Photonic Technology in Jena. A reflective layer ($200$ nm, $Ag$) in a predefined shape (axis of ellipse: 5.50 mm and 5.58 mm) was coated onto a glass substrate ($\o 1$ inch, 5 mm thick, $\lambda/10$ nominal surface flatness, BK-7, WG11050-A, Thorlabs, New Jersey, USA) and covered with a protective layer of $Al_2O_3$ ($65$ nm). The specific values for the layer thickness was found using a customized simulation, which calculated the reflectivity of stratified media using the \emph{matrix-method} \cite{Katsidis_MatrixMethod}, that predicted $\approx 95 \%$ reflectivity at $660$ nm for the given layer thicknesses. Note that this mirror has previously been used to separate super- and undercritical angle fluorescence, to achieve an improved precision in three-dimensional localization microscopy \cite{Dasgupta_Supercritical}.

\subsection{Experimental setup and evaluation process}
\label{sec:Experimental_setup}

\begin{figure}[htb]
	\centering
	\includegraphics[width=\linewidth]{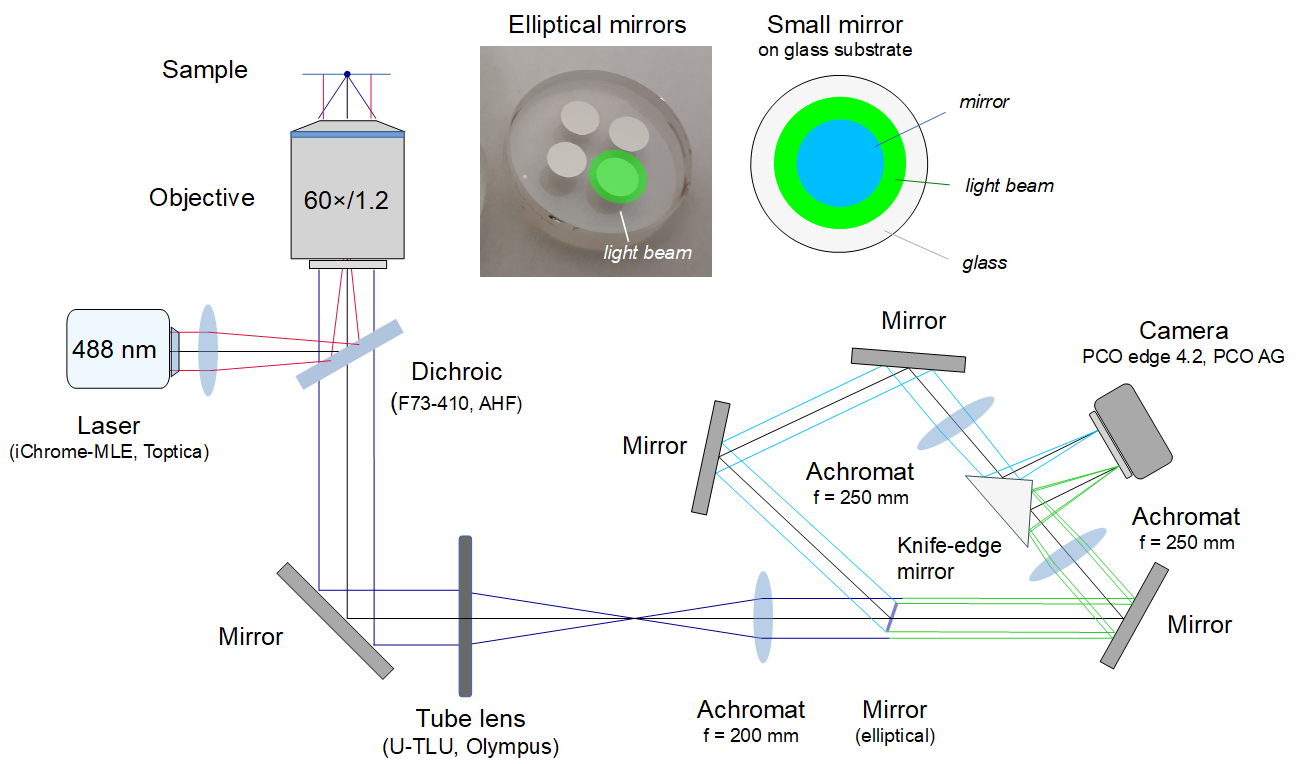}
	\caption{Experimental setup for achieving pupil splitting. The BFP of a microscope objective is being accessed through an additional lens ($f=200$ mm). A small mirror reflects the inner part of the pupil $\mathcal{\tilde{P}}_\Disk$, and the corresponding image is formed on one half of a camera. The image information passing through the annular pupil $\mathcal{\tilde{P}}_\Ring$ forms a second image on the remaining half of the same detector, enabling simultaneous acquisitions of the split image data. A more detailed description of the splitting mirror and the experimental setup can be found in Methods \ref{sec:Custom_EllipticalMirror} and \ref{sec:Experimental_setup}.}
	\label{fig:Fig5}
\end{figure}

The optical setup is shown in Fig. \ref{fig:Fig5} and has been described in detail in \cite{Dasgupta_Supercritical}. A collimated laser beam (488 nm, iChromeMLE, Toptica, Munich Germany) is focused at the BFP of the microscope objective ($60\times/1.2$, UPLSAPO60XW, Olympus, Tokyo, Japan), reflected by dichromatic beam splitter (F73-410, AHF, Tübingen, Germany). The detected fluorescence is collected by the objective, passes through the dichromatic beamsplitter and forms an image after the tube lens (U-TLU, Olympus, Tokyo, Japan). To access the BFP of the detection objective an achromatic lens ($f=200$ mm, Thorlabs, New Jersey, USA) was placed one focal length away from the primary image. The splitting mirror was mounted on a translation mount (Thorlabs, New Jersey, USA) and placed in the BFP of that lens. Both, the reflected and transmitted beams are imaged onto a camera (PCO edge 4.2, PCO AG, Kehlheim, Germany) using another achromatic lens ($f=250$ mm, Thorlabs, New Jersey, USA). The Argolight sample \cite{Royon_Argolight} (Argo-SIM, Argolight SA, Pessac, France) was positioned in the focal plane of the microscope, without the splitting mirror in place. The focus position was locked using a $785$ nm laser (iBeamSmart, Toptica, Munich, Germany) that was reflected by total-internal-reflection at the Argolight slide and then captured by a quadrant-diode (QP50-6 TO, First Sensor, Berlin, Germany). Once the resolution (pattern E) or stair target (pattern I) was found, a time series of widefield images (100 images) were taken. The laser intensity and exposure time were fixed to values such that the resulting images were dominated by shot noise. Without moving/refocusing the sample, changing the laser power or camera exposure, the splitting mirror was inserted and another set of 100 images was acquired. Note that those camera images contain both, the information corresponding to $\mathcal{\tilde{P}}_{\Disk}$ and $\mathcal{\tilde{P}}_{\Ring}$. All images were cropped manually to fit the respective image target and co-registered by finding a vertical/horizontal offset using an iterative cross-correlation approach \cite{Fienup_ImageRegistration}. As a last step background images were taken by blocking the light directly in front of the camera. Those background images were used to remove spatially varying offsets of the detector and to adjust the background level to zero average. Necessary information for the image recombination are the sub-PSFs $h_{1,2}$ of the system. Instead of imaging a bead sample to experimentally acquire the $h_{1,2}$, we chose to theoretically calculate the corresponding quantities, as shown in Methods \ref{sec:Computing_OTFs}. This indicates that pupil splitting does not rely on a very precise calibration of the system as the SNR enhancement was achievable without such a calibration.

\subsection{Calculating the visibility of the line patterns}
\label{sec:Calculating_visibility}

When a periodic signal is being imaged, the resulting modulation depth or visibility $\mathcal{V}$ is a direct indicator of the SNR performance (assuming proper noise-normalization). The visibility for such a signal can be computed according to \cite{Stelzer_SNR}:
\begin{equation}
	\mathcal{V} = \dfrac{I_{\Max} - I_{\Min}}{I_{\Max} + I_{\Min}}
\end{equation}
with $I_{\Max,\Min}$ being the maximum or minimum of the imaged signal. In case of a sinusoidal signal, $I_{\Max,\Min}$ corresponds to the maximum/minimum within one period. Note that when the intensity signal is raised by an amount $\varepsilon$, and the modulation depth of the curve stays the same, the aforementioned computation gives an erroneous visibility value $\mathcal{V}_{\varepsilon}$:
\begin{equation}
	\mathcal{V}_\varepsilon = \dfrac{ (I_{\Max}+\varepsilon) - (I_{\Min}+\varepsilon)}{(I_{\Max}+\varepsilon) + (I_{\Min}+\varepsilon)} = \dfrac{I_{\Max} - I_{\Min}}{I_{\Max} + I_{\Min} + 2 \cdot \varepsilon }
\end{equation}
So when comparing two different modulations which differ in their respective offset by $\varepsilon$, it is important to correct one of the calculated visibility values according to:
\begin{equation}
	\mathcal{V} = \mathcal{V}_\varepsilon \cdot \dfrac{I_{\Max} - I_{\Min}}{I_{\Max} - I_{\Min} - 2 \cdot \varepsilon \cdot \mathcal{V}_\varepsilon }
\end{equation}
In case of the data shown in Fig. \ref{fig:Fig4}, $\varepsilon$ was taken to be the average between the two maxima and the minimum of each image line pair (indicated with the circled numbers). 

}
\section{Bibliography}
\bibliographystyle{unsrtnat}
\bibliography{References}

\begin{acknowledgements}
We thank \emph{Polina Feldmann} for early discussions on the pupil splitting concept. 
\end{acknowledgements}

\begin{contributions}
R.H. conceived the initial idea. J.B. worked on the theoretical details and the numerical investigation. R.F. helped in the theoretical understanding. J.B. and A.J. designed the customized splitting mirror. U.H. manufactured the splitting mirror. J.B., T.D. and J.R. acquired the experimental data. J.B. analyzed the data and discussed with R.H. J.B. and R.H. wrote the first  draft, all authors contributed to and approved the final manuscript.
\end{contributions}


\begin{interests}
The authors declare no competing financial interests.
\end{interests}



\onecolumn
\newgeometry{left=63pt,right=53pt,bottom=63pt,top=53pt}

\section*{\Large{Supplementary information:}}
\label{sec:Supplement}

\subsection{Focusing light using Fermat's principle}
\label{sec:Fermat's principle}

\begin{figure}[htb]
	\centering
	\includegraphics[width=0.8\linewidth]{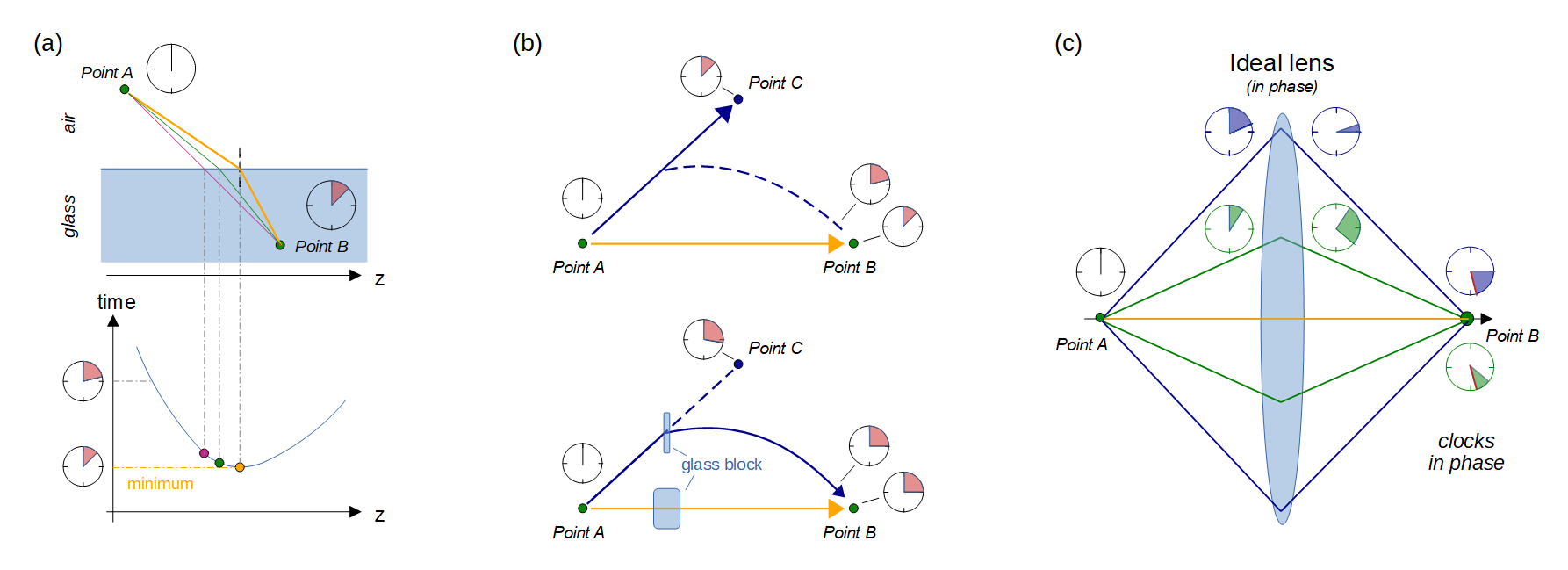}
	\caption{\textcolor{blue}{(a)} Using Fermat's principle to understand \emph{refraction} of light at an air/glass interface. Light always takes the path corresponding to the minimum amount of travel time (orange). \textcolor{blue}{(b)} When focusing all the light emitted by a point source (A) into another point (B), two glass block with different widths needs to be added so that the curved (dashed) and straight (orange) trajectory correspond to the same amount of time traveled, fulfilling Fermat's principle for those two possible light paths. \textcolor{blue}{(c)} Taking this "design" idea one step further, leads to the shape of the ideal lens. The time that light, emitted under an angle, loses due to the longer traveling distance to the lens surface, is compensated by reducing the propagation within the lens material, where light travels slower. Making sure Fermat's principle is fulfilled for all light rays emitted within the acceptance cone of the lens and yielding the highest possible concentration of light in B. Any deviation from the ideal lens shape will decrease the intensity arriving at the nominal focus, which makes a lens seemingly optimal.}
	\label{fig:FigA1}
\end{figure}

The shape of an ideal lens can be derived from \emph{Fermat's principle} of least time (or least optical path) \cite{Mahoney_Fermat}, as shown in \cite{Feynman_Lectures1}. This principle simply states that light takes the path corresponding to the shortest time it needs to travel. How light knows which of all possible paths corresponds to the one with the shortest time is not of interest here. Fermat's principle just states a simple rule, which already is able to describe many optical phenomena, e.g. refraction at an air/glass interface (see Fig. \ref{fig:FigA1} \textcolor{blue}{(a)}).

When light is supposed to travel between point \emph{A} and \emph{B}, which are separated by an air/glass interface, it takes the route of minimum optical path length (orange). Even though this requires the light to make a change of propagation direction, when entering the glass block.\\
The same argument can be made when thinking about how to focus light from a point source \emph{A} into point \emph{B}. Light is emitted in all directions and the challenge is to ensure that most of it ends up back at the nominal focus point B. 
When aiming to fulfill Fermat's principle, it needs to be ensured that the light traveling on a straight line from A to B is delayed with respect to the light traveling from A in the direction of C to B, e.g. by introducing two glass blocks with different width (Fig. \ref{fig:FigA1} \textcolor{blue}{(b)}). In this way we can make sure that both trajectories, straight (orange) and curved (blue), take the same optical path length. And because the path going diagonally through the smaller glass box (to reach C) is effectively longer than the curved version, the light emitted under an angle (towards C) will also end up in B. So, adding a glass block which gets thinner the further away from the line AB, will ensure that all the light emitted in a given light cone is focused into point B. This specifically shaped glass "block" is called an ideal \emph{lens} (Fig. \ref{fig:FigA1} \textcolor{blue}{(c)}). Light traveling under an angle will take longer to reach the lens surface, the time it has now "lost" is compensated by a shorter propagation distance inside the lens material (typically glass). Overall, this ensures that all light rays going to take exactly the same amount of time to travel from A to B, hence an ideal lens fulfills Fermat's principle. Furthermore it also shows that a maximum concentration of light, emitted by a point source, is reached using such an ideal lens for focusing. Any deviation from the ideal shape would violate Fermat's principle and cause a reduction of the intensity in its focal point.
	
\subsection{Incoherent image formation in real \& Fourier space}
\label{sec:IncoherentImageFormation}	

In incoherent imaging, the transfer of signal information in real space (denoted by cartesian coordinates $\vec{r} = [x,y]^\top$, with $^\top$ being the vector/matrix transpose) is given as a convolution operation $\otimes$ according to \cite{Goodman_Fourier}:
\begin{equation}
		I(\vec{r}) = S(\vec{r}) \otimes h(\vec{r}) 
		\label{eq:ImageFormation_real}
\end{equation}
with $I$ being the captured image, $S$ the unknown sample information and $h$ the respective PSF. The convolution $\otimes$ is defined as:
\begin{equation}
	I(\vec{r}) = \int_{-\infty}^{+\infty} d\vec{r}\hspace{2pt}^\prime \hspace{4pt} S(\vec{r}\hspace{2pt}^\prime) \cdot h(\vec{r}-\vec{r}\hspace{2pt}^\prime)
\end{equation}
In Fourier space (denoted by a tilde-sign) this relationship translates into a multiplication of the two respective spectra:
\begin{equation}
	\tilde{I}(\vec{k}) = \tilde{S}(\vec{k}) \cdot \tilde{h}(\vec{k})
	\label{eq:ImageFormation_Fourier}
\end{equation}
where $\tilde{h}$ describes the optical-transfer-function (OTF), $\tilde{S}$ the sample, $\tilde{I}$  the image information in Fourier space and $\vec{k}$ the spatial frequencies. The forward and inverse Fourier transform, denoted by $\text{  } \tilde{}$, is defined as \cite{Bronstein_Maths}:
\begin{equation}
	\tilde{f}(\vec{k}) = \mathcal{F} \big\{ f(\vec{r}) \big\} = \dfrac{1}{(2\pi)^{\Dim/2}} \int_{-\infty}^{+\infty} d\vec{r} \hspace{4pt} f(\vec{r}) \cdot e^{i \hspace{2pt} \vec{k} \cdot \vec{r}} \hspace{1cm} 	f(\vec{r}) = \mathcal{F}^{-1} \big\{ \tilde{f}(\vec{k}) \big\} = \dfrac{1}{(2\pi)^{\Dim/2}} \int_{-\infty}^{+\infty} d\vec{k} \hspace{4pt} \tilde{f}(\vec{k}) \cdot e^{-i \hspace{2pt} \vec{k} \cdot \vec{r}}
\end{equation}
with $f$/$\tilde{f}$ being the respective real/Fourier space quantity, $\vec{k} \cdot \vec{r}$ the scalar product and $\Dim$ the number of dimension of $\vec{r}$. 

\subsection{OTF of an aberrated pupil function}
\label{sec:Aberrated_Pupil}
The following analysis is based on chapter 6.4 \textit{Aberrations and their effects on frequency response} in \cite{Goodman_Fourier}.\\
The optical-transfer-function (OTF) is given as the autocorrelation of the pupil $\mathcal{\tilde{P}}$:
\begin{equation}
\tilde{h}(\vec{k}) = \mathcal{A} \big \{ \mathcal{\tilde{P}}(\vec{k}) \big \} = \int_{-\infty}^{+\infty} d\vec{k^\prime} \hspace{4pt} \mathcal{\tilde{P}}(\vec{k^\prime}) \cdot  \mathcal{\tilde{P}}^*(\vec{k^\prime} - \vec{k})
\end{equation}
where $^*$ represents complex-conjugation. Let's assume we have an aberrated or \emph{modified} pupil $\mathcal{\tilde{P}}_{\Mod}$ given as:
\begin{equation}
\mathcal{\tilde{P}}_{\Mod}(\vec{k}) = T(\vec{k}) \cdot e^{i \cdot  \varphi(\vec{k})} \cdot \mathcal{\tilde{P}}(\vec{k})
\end{equation}
with $\varphi \in \mathbb{R}$ being any phase aberration and $T \in \mathbb{R}$ a change in the amplitude transmission ($ 0 \leq T \leq 1$)  of $\mathcal{\tilde{P}}_{\Mod}$ .\\
The OTF corresponding to $\mathcal{\tilde{P}}_{\Mod}$ can now be expressed as:
\begin{equation}
\tilde{h}_{\Mod}(\vec{k}) = \int_{-\infty}^{+\infty} d\vec{k^\prime} \hspace{6pt}  T(\vec{k^\prime}) e^{i \cdot \varphi(\vec{k^\prime})} \mathcal{\tilde{P}}(\vec{k^\prime}) \hspace{4pt} \cdot \hspace{4pt} T(\vec{k^\prime}-\vec{k}) e^{-i \cdot \varphi(\vec{k^\prime}-\vec{k})} \mathcal{\tilde{P}}^*(\vec{k^\prime}-\vec{k})
\end{equation}
The absolute square of both, $\tilde{h}$ and $\tilde{h}_{\Mod}$, are given according to:
\begin{eqnarray}
|\tilde{h}(\vec{k})|^2 &=& \bigg| \int_{-\infty}^{+\infty} d\vec{k^\prime} \hspace{4pt} \mathcal{\tilde{P}}(\vec{k^\prime}) \cdot \mathcal{\tilde{P}}^*(\vec{k^\prime}-\vec{k}) \bigg|^2 \nonumber \\
|\tilde{h}_{\Mod}(\vec{k})|^2 &=& \bigg| \int_{-\infty}^{+\infty} d\vec{k^\prime} \hspace{4pt}  T(\vec{k^\prime}) e^{i \cdot \varphi(\vec{k^\prime})}  \mathcal{\tilde{P}}(\vec{k^\prime}) \hspace{4pt} \cdot \hspace{4pt} T(\vec{k^\prime}-\vec{k}) e^{-i \cdot \varphi(\vec{k^\prime}-\vec{k})} \mathcal{\tilde{P}}^*(\vec{k^\prime}-\vec{k})  \bigg|^2
\end{eqnarray}
Lets introduce the following substitution, to make the integrands more symmetric:
\begin{eqnarray}
\vec{k^\prime} - \vec{k} &=& \vec{u} - \vec{k}/2 \nonumber \\
\vec{k^\prime}  &=& \vec{u} + \vec{k}/2 \nonumber \\
d \vec{k^\prime} &=& d \vec{u}
\end{eqnarray}
And rewrite the previous equations with this change of variables into:
\begin{eqnarray}
|\tilde{h}(\vec{k})|^2 &=& \bigg| \int_{-\infty}^{+\infty} d\vec{u} \hspace{4pt} \mathcal{\tilde{P}}(\vec{u} + \vec{k}/2) \cdot \mathcal{\tilde{P}}^*(\vec{u}-\vec{k}/2) \bigg|^2 \nonumber \\
|\tilde{h}_{\Mod}(\vec{k})|^2 &=& \bigg| \int_{-\infty}^{+\infty} d\vec{u} \hspace{4pt} T(\vec{u} + \vec{k}/2) e^{i \cdot \varphi(\vec{u} + \vec{k}/2)} \mathcal{\tilde{P}}(\vec{u} + \vec{k}/2) \hspace{4pt} \cdot \hspace{4pt} T(\vec{u}-\vec{k}/2)  e^{-i \cdot \varphi(\vec{u}-\vec{k}/2)}  \mathcal{\tilde{P}}^*(\vec{u}-\vec{k}/2)  \bigg|^2
\end{eqnarray}
Using the \emph{inequality of Schwartz} \cite{Bronstein_Maths} on $|\tilde{h}_{\Mod}|^2$, we find that:
\begin{equation}
|\tilde{h}_{\Mod}(\vec{k})|^2 \leq  \bigg[\int_{-\infty}^{+\infty} d\vec{u} \hspace{4pt} | T(\vec{u} + \vec{k}/2)|^2 \cdot | \mathcal{\tilde{P}}(\vec{u} + \vec{k}/2)|^2 \bigg] \cdot  \bigg[\int_{-\infty}^{+\infty} d\vec{u} \hspace{4pt} | T(\vec{u} - \vec{k}/2)|^2 \cdot | \mathcal{\tilde{P}}(\vec{u} - \vec{k}/2)|^2 \bigg]
\end{equation}
Which shows that the largest value of $\tilde{h}_{\Mod}$ is given for $T = 1$ (i.e. constant amplitude transmission) and yields in:

\if
With this we can simplify $|\tilde{h}_{\Mod}|^2$ as both brackets show the same integral:
\begin{equation}
|\tilde{h}_{\Mod}(\vec{k})|^2 \leq  \bigg[\int_{-\infty}^{+\infty} d\vec{u} \hspace{4pt} | T(\vec{u} - \vec{k}/2)|^2 \cdot | \mathcal{\tilde{P}}(\vec{u} - \vec{k}/2)|^2 \bigg] ^2
\end{equation}
Accordingly, the expression of $|\tilde{h}|^2$ is given as:
\begin{eqnarray}
|\tilde{h}(\vec{k})|^2 &=& \bigg| \int_{-\infty}^{+\infty} d\vec{u} \hspace{4pt} \mathcal{\tilde{P}}(\vec{u} - \vec{k}/2) \cdot \mathcal{\tilde{P}}^*(\vec{u}-\vec{k}/2) \bigg|^2 = \nonumber \\
&=& \bigg| \int_{-\infty}^{+\infty} d\vec{u} \hspace{4pt} |\mathcal{\tilde{P}}(\vec{u} - \vec{k}/2)|^2 \bigg|^2 \geq \bigg[\int_{-\infty}^{+\infty} d\vec{u} \hspace{4pt} | \textcolor{blue}{T(\vec{u} - \vec{k}/2)|^2} \cdot | \mathcal{\tilde{P}}(\vec{u} - \vec{k}/2)|^2 \bigg] ^2
\end{eqnarray}
Showing that even if we would neglect any modification of the transmission of $\mathcal{\tilde{P}}_{\Mod}$ and only allow phase aberrations, the absolute square of the corresponding OTF is always smaller or equal to the non-aberrated pupil $\mathcal{\tilde{P}}$:
\fi
\begin{equation}
\boxed{|\tilde{h}_{\Mod}(\vec{k})|^2 \leq |\tilde{h}(\vec{k})|^2}
\end{equation}
Hence, it is impossible to improve the signal transfer characteristics of a lens, by any kind of amplitude or phase modulation in the pupil. 

\subsection{Shot noise in real \& Fourier space}	
\label{sec:Photon-limited noise}
In noisy images the information in a single measurement is not directly given through eq. \ref{eq:ImageFormation_real} or \ref{eq:ImageFormation_Fourier}. Instead, it is expressed by an expectation value $\mu$ \& $\tilde{\mu}$ and noise ($\mathcal{N}$ \& $\tilde{\mathcal{N}}$), which makes the measurement deviate from the expectancy:
\begin{eqnarray}
	I(\vec{r}) &=& \mu(\vec{r}) + \mathcal{N}(\vec{r}) = S(\vec{r}) \otimes h(\vec{r}) + \mathcal{N}(\vec{r}) \label{eq:Mu_realSpace}\\
	\tilde{I}(\vec{k}) &=&  \tilde{\mu}(\vec{k}) + \mathcal{\tilde{N}}(\vec{k}) = \tilde{S}(\vec{k}) \cdot \tilde{h}(\vec{k}) + \tilde{N}(\vec{k}) \label{eq:Mu_FourierSpace}
\end{eqnarray}
Shot noise is Poisson-distributed, which relates the variance $\sigma^2$ of the recorded images $I$ directly to its expectation value $\mu$ via \cite{Lucke_PoissonNoise}: 
\begin{equation}
	\boxed{\Var\{ I(\vec{r}) \} = \sigma^2(\vec{r}) = \mu(\vec{r})}
\end{equation}
The additive noise component in real space is bias-free ($\langle \mathcal{N} \rangle = 0$) and scales with the expectation value of $I$ ($\Var\{\mathcal{N}\} = \mu$).\\
The expectation value of the noise in Fourier space can be computed as:
\begin{eqnarray}
	\langle \tilde{\mathcal{N}} (\vec{k}) \rangle &=& \bigg \langle \mathcal{F}\{ \mathcal{N}(\vec{r}) \} \bigg \rangle =  \bigg \langle \dfrac{1}{(2\pi)^{\Dim/2}} \int_{-\infty}^{+\infty} d\vec{r} \hspace{4pt} \mathcal{N}(\vec{r}) \cdot e^{i \hspace{2pt} \vec{k} \cdot \vec{r}} \bigg \rangle = \nonumber \\  
	&=& \dfrac{1}{(2\pi)^{\Dim/2}} \int_{-\infty}^{+\infty} d\vec{r} \hspace{4pt} \underbrace{\langle \mathcal{N}(\vec{r}) \rangle}_{=0} \cdot e^{i \hspace{2pt} \vec{k} \cdot \vec{r}} = 0
\end{eqnarray}
with $\mathcal{F}$ denoting a Fourier transform. Meaning that the noise is also bias-free ($\langle \tilde{\mathcal{N}} \rangle = 0$) in Fourier space. Because of this, the corresponding variance can simply be expressed as: 
\begin{eqnarray}
	\Var \{ \tilde{\mathcal{N}} (\vec{k}) \} &=& \langle |\tilde{\mathcal{N}} (\vec{k}) |^2 \rangle - |\underbrace{\langle  \tilde{\mathcal{N}} (\vec{k}) \rangle}_{=0}|^2 = \langle \tilde{\mathcal{N}}(\vec{k}) \cdot \tilde{\mathcal{N}}^*(\vec{k}) \rangle = \nonumber \\
	&=&\big \langle \mathcal{F} \bigg\{ \mathcal{N}(\vec{r}) \otimes \mathcal{N}(-\vec{r}) \bigg\} \big \rangle
\end{eqnarray}
With $\otimes$ denoting a convolution operation that, when given in its integral form, can be used to express the argument of the aforementioned expectation value as:
\begin{eqnarray}
	\mathcal{F} \bigg\{ \mathcal{N}(\vec{r}) \otimes \mathcal{N}(-\vec{r}) \bigg\} &=& \dfrac{1}{(2\pi)^{\Dim/2}} \int_{-\infty}^{+\infty} d\vec{r} \hspace{4pt}  \left[\mathcal{N}(\vec{r}) \otimes \mathcal{N}(-\vec{r}) \right] \cdot e^{i \hspace{2pt} \vec{k} \cdot \vec{r}} = \nonumber \\
	&=&\dfrac{1}{(2\pi)^{\Dim/2}} \int_{-\infty}^{+\infty} d\vec{r} \hspace{4pt} e^{i \hspace{2pt} \vec{k} \cdot \vec{r}} \int_{-\infty}^{+\infty} d\vec{r^\prime} \hspace{4pt} \mathcal{N}(\vec{r^\prime}) \cdot \mathcal{N}(-\vec{r} + \vec{r^\prime})
\end{eqnarray}
We can shift the expectation value into the integral computation, yielding:
\begin{equation}
	\Var\{\tilde{\mathcal{N}}(\vec{k})\} = \dfrac{1}{(2\pi)^{\Dim/2}} \int_{-\infty}^{+\infty} d\vec{r} \hspace{4pt} e^{i \hspace{2pt} \vec{k} \cdot \vec{r}} \int_{-\infty}^{+\infty} d\vec{r^\prime} \hspace{4pt} \langle \mathcal{N}(\vec{r^\prime}) \cdot \mathcal{N}(-\vec{r} + \vec{r^\prime}) \rangle
\end{equation}

The term in the second integral is the expectancy of measuring noise at two different positions, namely, $\vec{r^\prime}$ and $-\vec{r} + \vec{r^\prime}$. We assume that the noise is \emph{pixel-independent} (true for shot noise), hence the aforementioned expectation value is only non-zero when $\vec{r} = 0$, yielding:
\begin{equation}
	\langle \mathcal{N}(\vec{r^\prime}) \cdot \mathcal{N}(-\vec{r} + \vec{r^\prime}) \rangle = \underbrace{\langle \mathcal{N}(\vec{r^\prime})^2 \rangle}_{= \Var\{\mathcal{N}(\vec{r^\prime})\}} \cdot \delta(\vec{r})  = \mu(\vec{r^\prime}) \cdot \delta(\vec{r}) 
\end{equation}
With this the variance of the noise component in Fourier space can be simplified to:
\begin{eqnarray}
	\Var\{\tilde{\mathcal{N}}(\vec{k})\} &=& \dfrac{1}{(2\pi)^{\Dim/2}} \int_{-\infty}^{+\infty} d\vec{r^\prime} \hspace{4pt} \mu(\vec{r^\prime}) \int_{-\infty}^{+\infty} d\vec{r} \hspace{4pt} e^{i \hspace{2pt} \vec{k} \cdot \vec{r}}  \cdot \delta(\vec{r}) = \nonumber \\
	&=&\dfrac{1}{(2\pi)^{\Dim/2}} \int_{-\infty}^{+\infty} d\vec{r^\prime} \hspace{4pt} \mu(\vec{r^\prime}) \cdot e^{i \hspace{2pt} \vec{k} \cdot \vec{0}}
\end{eqnarray}
Yielding the result that $\Var\{\tilde{\mathcal{N}}\}$ is independent of $\vec{k}$ and proportional to the total number of detectable photons $p$.
\begin{equation}
	\Var \{ \tilde{\mathcal{N}}(\vec{k}) \} \propto \int_{-\infty}^{+\infty} d\vec{r} \hspace{4pt} \mu(\vec{r}) = p
	\label{eq:total_number_of_photons}
\end{equation}
Because we know that $\tilde{\mathcal{N}}$ is bias-free, we conclude that the variance of $\tilde{I}$ is also constant in Fourier space and proportional to $p$:
\begin{equation}
	\boxed{\Var \{ \tilde{I}(\vec{k}) \} \propto  p}
\end{equation}
\if
Detecting photons inevitably will lead to shot noise. Meaning that the values of $I$ in each pixel are \emph{Poisson} distributed \cite{Lucke_PoissonNoise}. Noise is typically characterized by the expectation value $\mu$ and the variance $\sigma^2$. In case of the Poisson distributed $I$ this yields:
\begin{equation}
	\mu (\vec{r}) = \sigma^2(\vec{r})
	\label{eq:Poisson_property}
\end{equation} 	
with the expectation value in real space given as:
\begin{equation}
	\mu(\vec{r})  = S(\vec{r}) \otimes h(\vec{r})
	\label{eq:expectancy_real_space}
\end{equation}
This shows that the noise is sample dependent, as it scales with the expectation value, e.g. in the trivial case where no photons are being expected, the measurement will be noise-free. The expectation value $\tilde{\mu}$ in Fourier space is given according to:
\begin{equation}
	\tilde{\mu}(\vec{k}) = \tilde{S}(\vec{k}) \cdot \tilde{h}(\vec{k})
\end{equation}
Lets denote $\tilde{\sigma}^2$ as the variance of the image $\tilde{I}$ in Fourier space. It is calculated by adding the variance of the real and imaginary part \cite{Bronstein_Maths}:
\begin{equation}
	\tilde{\sigma}^2 (\vec{k}) = \Var \big [\operatorname{Re} \{ \tilde{I}(\vec{k}) \} \big ] + \Var \big[ \operatorname{Im} \{ \tilde{I}(\vec{k}) \} \big] \hspace{4pt} \in \hspace{4pt} \mathbb{R} 
\end{equation}
with $\Var$ representing the variance operation and $\operatorname{Re}/\operatorname{Im}$ the real/imaginary part. We can derive $\tilde{\sigma}^2$ by Fourier transforming the image $I$:
\begin{equation}
	\tilde{\sigma}^2 (\vec{k}) = Var \big [ \tilde{I}(\vec{k}) \big ] = \Var \big [ \mathcal{F} \{ I(\vec{r}) \} \big] = \Var \left[ \dfrac{1}{(2\pi)^{\Dim/2}} \int_{-\infty}^{+\infty} d\vec{r} \hspace{4pt} I(\vec{r}) \cdot e^{i \hspace{2pt} \vec{k} \cdot \vec{r}} \right]
\end{equation}
We exchange the $\Var$- and the $\mathcal{F}$-operator (due to linearity) and assume that the noise is statistically independent with respect to $\vec{r}$ \cite{Lucke_PoissonNoise}:
\begin{equation}
	\tilde{\sigma}^2 (\vec{k}) = \dfrac{1}{(2\pi)^{\Dim/2}} \int_{-\infty}^{+\infty} d\vec{r} \hspace{4pt}  \Var \big[ I(\vec{r}) \cdot e^{i \hspace{2pt} \vec{k} \cdot \vec{r}} \big]
\end{equation}
In terms of calculating the variance the complex exponential is $=1$, hence we find:
\begin{equation}
	\tilde{\sigma}^2 (\vec{k}) = \dfrac{1}{(2\pi)^{\Dim/2}} \int_{-\infty}^{+\infty} d\vec{r} \hspace{4pt} \left|e^{i \hspace{2pt} \vec{k} \cdot \vec{r}} \right|^2 \cdot  \Var \big[ I(\vec{r})\big] = \dfrac{1}{(2\pi)^{\Dim/2}} \int_{-\infty}^{+\infty} d\vec{r} \hspace{4pt} \Var \big[ I(\vec{r})\big] 
\end{equation}
The variance of $I$ is equal to the expectation value $\mu$ (see eq. \ref{eq:Poisson_property}). This means that the variance in Fourier space is proportional to the integral of all expected values in real space, the photon budget $p$:
\begin{equation}
 	\tilde{\sigma}^2 (\vec{k})  = \dfrac{1}{(2\pi)^{\Dim/2}} \int_{-\infty}^{+\infty} d\vec{r} \hspace{4pt} \hspace{2pt} \mu(\vec{r}) \hspace{4pt} \propto \hspace{4pt} p
\end{equation}
Note that the variance in Fourier space does not depend on $\vec{k}$ and scales with the amount of photons present in $I$.\\
We can approximate the value of $\tilde{\sigma}^2$ directly from the Fourier transform $\tilde{I}$ as the $\vec{k}=\vec{0}$ - component:
\begin{equation}
	\tilde{I}(\vec{k}=\vec{0}) = \dfrac{1}{(2\pi)^{\Dim/2}} \int_{-\infty}^{+\infty} d\vec{r} \hspace{4pt} I(\vec{r}) \cdot e^{i \hspace{2pt} \vec{0} \cdot \vec{r}}  = \dfrac{1}{(2\pi)^{\Dim/2}} \int_{-\infty}^{+\infty} d\vec{r} \hspace{4pt} I(\vec{r}) \approx \dfrac{1}{(2\pi)^{\Dim/2}} \int_{-\infty}^{+\infty} d\vec{r} \hspace{4pt} \mu(\vec{r})
\end{equation}
\fi

\subsection{Signal-to-noise ratio in real \& Fourier space}
\label{sec:SNR_Real_Fourier}

The SNR is conveniently described as the measure $\SNR = \mu / \sigma$ \cite{Stelzer_SNR}.  For shot noise, in real space, this translates into:
\begin{equation}
	\SNR(\vec{r}) =  \dfrac{\mu(\vec{r})}{\sqrt{\mu(\vec{r})}} =  \sqrt{\mu(\vec{r})}
\end{equation}
Indicating that the SNR is proportional to the square root of the expectation value which can be increased by manipulating either the sample emission $S$ or the detection PSF $h$ (see equation \ref{eq:Mu_realSpace}). For a fixed photon budget $p$ the sample emission is fixed and, according to Fermat's principle, $h$ is maximal at the focus where it also has the best SNR. The SNR can seemingly only be improved by capturing more photons.\\
In Fourier space the same general definition for the SNR can be used, while using the absolute value to ensure $\SNR(\vec{k}) \in \mathbb{R}$:
\begin{equation}
	\SNR(\vec{k}) = \dfrac{| \tilde{\mu}(\vec{k})|}{\sqrt{p}} = \sqrt{p} \cdot |\tilde{\mu}_0(\vec{k})|
	\label{eq:SNR_Fourier}
\end{equation}
with $\tilde{\mu}_0 = |\tilde{\mu}| / p$ being a normalized expectation value ($\tilde{\mu}_0 (\vec{0}) = 1$) and $p$ the total number of expected photons (see eq. \ref{eq:total_number_of_photons}).\\
This makes clear that increasing the available photon budget will improve the attainable SNR. As $\tilde{\mu}_0$ for most natural objects is a monotonically decreasing function \cite{Hsiao_NaturalObjects} (e.g. an exemption would be a cosine-grating), it can be seen that the SNR drops for larger spatial frequencies $\vec{k}$ and vanishes at the cut-off frequency $k_{\WF}$.  Meaning that a reduction of the photon budget $p$ will first come at the cost of high spatial frequency information and therefore limit the achievable spatial resolution.  

\subsection{OTF as an autocorrelation of the full \& ring pupil}
\label{sec:OTF_autocorrelation}

The OTF of an imaging system is connected to its pupil via an autocorrelation operation $\mathcal{A}$ \cite{Goodman_Fourier}, defined as:
\begin{equation}
	\tilde{h}(\vec{k}) = \mathcal{A} \big \{ \mathcal{\tilde{P}}(\vec{k}) \big \} = \int_{-\infty}^{+\infty} d\vec{k^\prime} \hspace{4pt} \mathcal{\tilde{P}}(\vec{k^\prime}) \cdot  \mathcal{\tilde{P}}(\vec{k^\prime} - \vec{k})
\end{equation}
where we already assumed that the pupil of an ideal system is real-valued only (no phase $\varphi$ \& amplitude $T$ aberrations), so that $\mathcal{\tilde{P}}^* = \mathcal{\tilde{P}}$.\\
The widefield pupil $\mathcal{\tilde{P}}_{\WF}$ is split into two components, fulfilling:
\begin{equation}
	\mathcal{\tilde{P}}_{\WF}(\vec{k}) = \mathcal{\tilde{P}}_{\Disk}(\vec{k}) + \mathcal{\tilde{P}}_{\Ring}(\vec{k})
\end{equation}
When the aforementioned autocorrelation integral is solved, the following relationship between the different OTFs is found:
\begin{equation}
	\tilde{h}_{\WF}(\vec{k}) = \tilde{h}_{\Disk}(\vec{k}) + \tilde{h}_{\Ring}(\vec{k}) + 2 \cdot \mathcal{\tilde{P}}_{\Disk}(\vec{k}) \otimes \mathcal{\tilde{P}}_{\Ring} (\vec{k})
\end{equation}
assuming the pupils to be real-valued transmissions of symmetric shape. With the last term resulting from the missing interference between light that is now split into the disk and ring pupil.
We introduce $k_0 = 0.5 \cdot (k_{\WF} + k_{\Disk})$, the average of the maximum transferable spatial frequency of the split and widefield pupil. Note that for $|\vec{k}| \geq k_0$ the interference between $\mathcal{\tilde{P}}_{\Disk}$ and $\mathcal{\tilde{P}}_{\Ring}$, as well as $\tilde{h}_{\Disk}$ vanishes, making the signal transfer in WF and ring pupil imaging equivalent:
\begin{equation}
	\tilde{h}_{\WF}\left(|\vec{k}| \geq k_0\right) = \tilde{h}_{\Ring} \left(|\vec{k}| \geq k_0\right)
\end{equation}
Proving that it is possible to transfer spatial frequencies for which $|\vec{k}| \geq k_0$, with the same signal strength by only using the ring pupil.\\

\begin{figure}[htb]
	\centering
	\begin{subfigure}[c]{0.45\textwidth}
		\includegraphics[width=\linewidth]{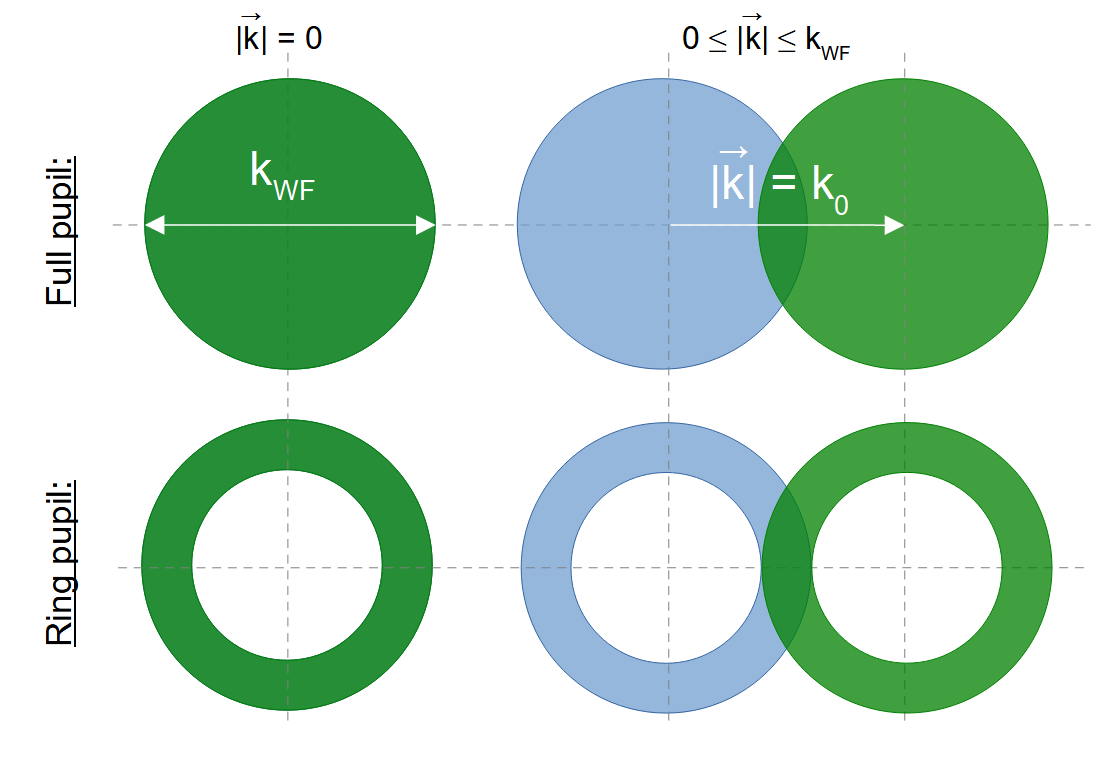}
		\caption{}
		\label{fig:FigA2a}
	\end{subfigure}
	\hspace{1cm}
	\begin{subfigure}[c]{0.45\textwidth}
		\includegraphics[width=\linewidth]{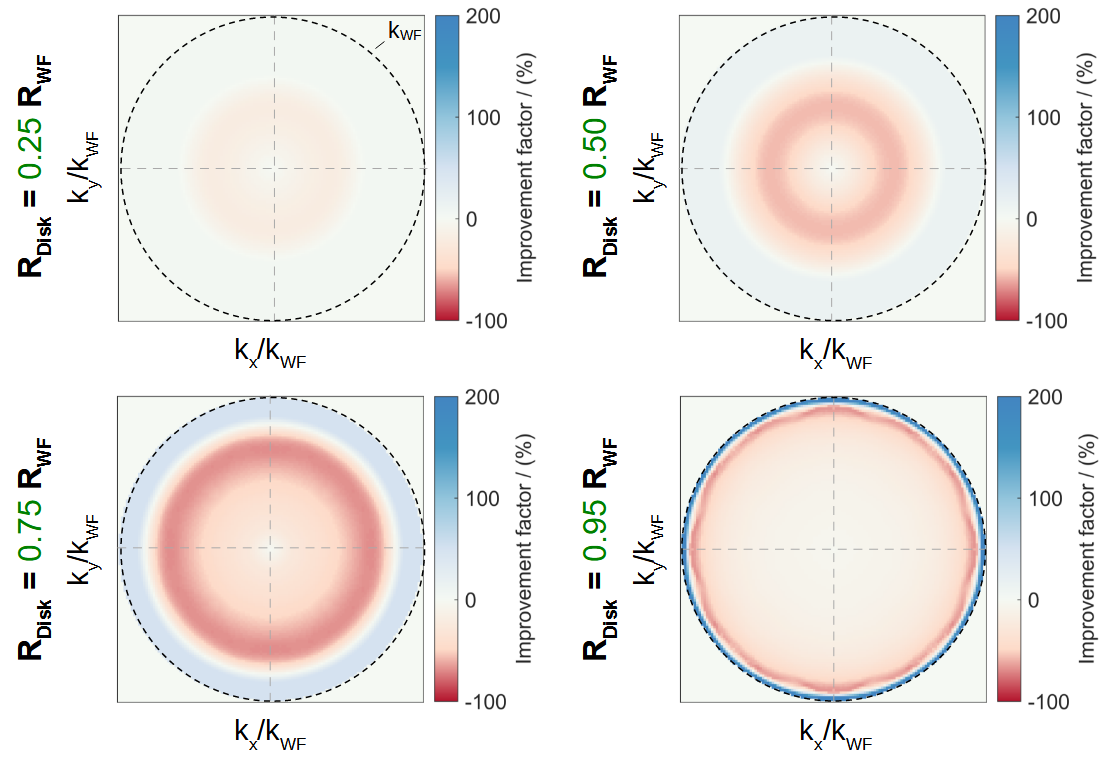}
		\caption{}
		\label{fig:FigA2b}
	\end{subfigure}
	\caption{\textcolor{blue}{(a)} Visualization of a geometrical way how to obtain the OTF value at $\vec{k}$, from a given pupil $\mathcal{\tilde{P}}$ (assuming no phase or amplitude modulation). Two copies of the corresponding pupil (top: WF, bottom: Ring) are centered on top of each other, and the resulting overlap is proportional to $|\tilde{h}(\vec{0})|$. The diameter of the pupil is given as the maximum transferable spatial frequency $k_{\WF}$. The two copies are shifted apart, the remaining overlap is monitored and the center to center distance of the shifted pupils is regarded as $|\vec{k}|/k_{\WF}$. Note that for a specific shift $|\vec{k}|=k_0$ the overlap is the same, no matter whether a full or ring pupil is used. \textcolor{blue}{(b)} Overview of the improvement factor $\IF$ with respect to changing the splitting parameter $\eta$ (green). Strong enhancement can only be reached at high spatial frequencies, by making the outer ring pupil very thin. Note the tradeoff between improvement strength $\IF_{\Max}$ and improvable region $\IF>0$. We have chosen $\eta = \sqrt{0.5}$, which yields as split into equal photon numbers.}
	\label{fig:FigA2}
\end{figure}

Another way to observe the aforementioned "same-signal-transfer" effect is by picturing the autocorrelation operation in a graphical way. For obtaining the OTF values corresponding to a specific pupil $\mathcal{\tilde{P}}$, we need to take two copies of $\mathcal{\tilde{P}}$, shift them relative to each other and calculate the overlap area of the newly formed structure. This process is shown in Fig. \ref{fig:FigA2a} for the widefield and ring pupil. 

From a geometrical viewpoint, it becomes clear now why the curve of |OTF| is decreasing for higher frequencies: the overlap (dark green) gets smaller the further we separate the two pupils. The region-of-support is reached exactly at two times the pupil radius $k_\WF$, which equates in the Abbe limit \cite{Abbe_Beitraege}. Using an annular pupil drastically reduces the overlap for most frequencies, yielding a reduced transfer strength at mid-frequencies. However, we can see from Fig. \ref{fig:FigA2a} that for the very high spatial frequencies $|\vec{k}| \geq k_0$, the overlap is equivalent to that of the WF case. This explains, why the green OTF-curve in Fig. \ref{fig:Fig2} has exactly the same values, for $|\vec{k}|\geq k_0$ as the blue curve.

\subsection{Weighted averaging in Fourier space}
\label{sec:WeightedAvg}

By employing pupil splitting we are able to measure two sub-images $I_{1,2}$, which correspond to the two sub-pupils $\tilde{\mathcal{P}}_{\Disk,\Ring}$. Our goal is now to recombine both in such a way that the resulting $\SNR(\vec{k})$ in Fourier space is maximal \cite{Wicker_PhDThesis,Becker_Better,Heintzmann_Subtraction}. This is achieved by computing the following weighted average $I_{wa}$:
\begin{equation}
	\tilde{I}_{wa}(\vec{k}) = \tilde{w}_1(\vec{k}) \cdot \tilde{I}_1(\vec{k}) + \tilde{w}_2(\vec{k}) \cdot \tilde{I}_2(\vec{k})
\end{equation}
with $\tilde{w}_{1,2}$ being the individual weights, which determine the importance of each image at a particular spatial frequency.\\
In the following we want to find an expression for those weights by maximizing the SNR in Fourier space. For this we need to calculate the expectation value and noise variance of the recombined image:
\begin{eqnarray}
	\tilde{\mu}_{wa} (\vec{k}) &=& \big[ \tilde{w}_1(\vec{k}) \cdot \tilde{h}_1(\vec{k}) + \tilde{w}_2(\vec{k}) \cdot \tilde{h}_2(\vec{k}) \big] \cdot \tilde{S}(\vec{k})\\
	\tilde{\sigma}_{wa}^2 (\vec{k}) &=& \tilde{w}_1^2 (\vec{k}) \cdot \tilde{\sigma}_1^2 + \tilde{w}_2^2 (\vec{k}) \cdot \tilde{\sigma}_2^2
\end{eqnarray}
We can now define the SNR of $\tilde{I}_{wa}(\vec{k})$ according to eq. \ref{eq:SNR_Fourier} as:
\begin{equation}
	\SNR_{wa} (\vec{k}) = \dfrac{|\tilde{\mu}_{wa} (\vec{k})|}{\tilde{\sigma}_{wa} (\vec{k})} =  \dfrac{ \tilde{w}_1(\vec{k}) \cdot \tilde{h}_1(\vec{k}) + \tilde{w}_2(\vec{k}) \cdot \tilde{h}_2(\vec{k})}{\sqrt{\tilde{w}_1^2 (\vec{k}) \cdot \tilde{\sigma}_1^2 + \tilde{w}_2^2 (\vec{k}) \cdot \tilde{\sigma}_2^2}} \cdot \tilde{S}(\vec{k})
\end{equation}
The goal is find the optimal $\tilde{w}_{1,2}$ that maximize the SNR: hence we need to look for the maximum of $\SNR_{wa}(\vec{k}) / \tilde{S}(\vec{k})$. The derivative with respect to the weight $\tilde{w}_l$ is found to be:
\begin{equation}
	\tfrac{1}{\tilde{S}(\vec{k})} \cdot \dfrac{\partial \SNR_{wa}(\vec{k}) }{\partial \tilde{w}_l(\vec{k})} = \dfrac{\tilde{h}_l(\vec{k}) \cdot \big[ \tilde{w}_1^2 (\vec{k}) \tilde{\sigma}_1^2 + \tilde{w}_2^2 (\vec{k}) \tilde{\sigma}_2^2 \big] - \tilde{w}_l(\vec{k}) \tilde{\sigma}_l^2 \cdot \big[ \tilde{w}_1(\vec{k}) \tilde{h}_1(\vec{k}) + \tilde{w}_2(\vec{k})  \tilde{h}_2(\vec{k})\big] }{\sqrt[3/2]{\tilde{w}_1^2 (\vec{k}) \cdot \tilde{\sigma}_1^2 + \tilde{w}_2^2 (\vec{k}) \cdot \tilde{\sigma}_2^2}}
\end{equation}
Setting this expression to zero yields a criterion which needs to be fulfilled when the weighted average is sought to maximize the SNR:
\begin{equation}
	\tilde{h}_l(\vec{k}) \cdot \big[ \tilde{w}_1^2 (\vec{k}) \tilde{\sigma}_1^2 + \tilde{w}_2^2 (\vec{k}) \tilde{\sigma}_2^2 \big] = \tilde{w}_l(\vec{k}) \tilde{\sigma}_l^2 \cdot \big[ \tilde{w}_1(\vec{k}) \tilde{h}_1(\vec{k}) + \tilde{w}_2(\vec{k})  \tilde{h}_2(\vec{k})\big]
	\label{eq:WeightedAvg_equality}
\end{equation}
Together with the additional conditions of noise-normalization and weights adding to unity (see \cite{Becker_PhD} for more details):
\begin{eqnarray}
	\tilde{\sigma}_{wa}^2 (\vec{k})  &=& 1\\
	\tilde{w}_1(\vec{k})  + \tilde{w}_2(\vec{k}) &=& 1
\end{eqnarray}

, yields in the following expression for the weights:
\begin{equation}
	\tilde{w}_l(\vec{k}) = \dfrac{\tilde{h}_l(\vec{k}) / \tilde{\sigma}_l^2}{\sqrt{\tilde{h}_1^2(\vec{k}) / \tilde{\sigma}_1^2 + \tilde{h}_2^2(\vec{k}) / \tilde{\sigma}_2^2}}
\end{equation}
This enables us to theoretically predict the performance of the pupil splitting \& weighted average recombination method by computing the improvement factor $\IF$ in the next section.

\subsection{Improvement factor for different radial splits}
\label{sec:ImprovementFactor}

We want to investigate how the SNR improvement at high spatial frequencies varies with changing radial splitting, given by $\eta$. To do this we define the improvement factor $\IF$, according to \cite{Becker_Better}:
\begin{equation}
	\IF(\vec{k}) = \dfrac{|\tilde{h}_{wa } (\vec{k})|}{|\tilde{h}_{\WF}(\vec{k})|} -1
\end{equation}
In case of no improvement $\IF = 0$, as then $\tilde{h}_{wa } = \tilde{h}_{\WF}$. An improvement is indicated by $\IF > 0$, a worsening by $\IF < 0$. Note that complete loss of information corresponds to $\tilde{h}_{wa } = 0$, which yields $\IF = -1$. The improvement factor is depicted in Fig. \ref{fig:FigA2b}, for four different radial splittings corresponding to $\eta = [0.25, \hspace{2pt} 0.5, \hspace{2pt} 0.75, \hspace{2pt} 0.95]$. We can see that $\IF$ becomes very large when $R_\Disk$ approaches the pupil radius $R_{\WF}$ (or $\eta \rightarrow 1$). However, the fraction of $\vec{k}$-space which can be enhanced gets more restricted to frequencies very close to the cut-off limit $k_{\WF}$, which means that a tradeoff needs to be achieved. In our work we selected $\eta = R_\Disk / R_{\WF} = \sqrt{0.5} \approx 0.707$, so that both sub-images carry the same number of photons. This will yield a maximum improvement value of $\approx 41 \%$.\\
Another important aspect to note is, that the "negative improvement" stays well away from $\IF = -1$. Which means that, although those frequencies may be transferred worse, we can still recover them by means of computational reconstruction (provided their are above the noise floor). 

\subsection{Multiview Richardson-Lucy deconvolution for image fusion}
\label{sec:Multiview_Deconvolution}

Previous work (e.g. \cite{Heintzmann_Multiview,York_MultiviewDeconvolution}) has shown how multiple images can be recombined using the Richardson-Lucy (RL) algorithm \cite{Richardson_Deconvolution,Lucy_Deconvolution}.\\
RL is typically used to deconvolve single images that are corrupted by \emph{Poissonian} noise. This is done by iteratively computing the most probable estimate $\hat{S}$ of the object distribution, which itself would result in the observed (2D) image $I$, when being blurred by the detection PSF. In each iteration the current estimate $\hat{S}$ is multiplied by a correction factor $C$, according to:
\begin{equation}
	\hat{S}_{j+1} (\vec{r}) = \hat{S}_j (\vec{r}) \cdot C_j(\vec{r}) = \hat{S}_j (\vec{r}) \cdot \bigg[ \dfrac{I(\vec{r})}{\hat{S}_j (\vec{r}) \otimes h(\vec{r})} \otimes h(-\vec{r}) \bigg]
\end{equation}
with $j$ being the current iteration number and $C$ given by the conventional RL scheme introduced in \cite{Lucy_Deconvolution,Richardson_Deconvolution}. When dealing with multiple images of the same object it is possible to extend the iterative algorithm so that a correction factor is obtained for all sub-images. As there is only a single object estimate, those correction factors need to be combined, e.g. by summation as done in \cite{York_MultiviewDeconvolution}:
\begin{equation}
	C_j(\vec{r}) =  \sum_{l=1}^2 \bigg[ \dfrac{I_l(\vec{r})}{\hat{S}_j (\vec{r}) \otimes h_l(\vec{r})} \otimes h_l(-\vec{r}) \bigg]
\end{equation}
Here we further propose to use the "weighting" idea, already used in sec. \ref{sec:WeightedAvg}, in conjunction with the RL reconstruction.
The idea is to weight the individual correction factors such that a modified iterative update is achieved. The \emph{weighted} correction factor are then given as:
\begin{equation}
	C_j(\vec{r}) =   \mathcal{F}^{-1} \Bigg\{\sum_{l=1}^2 \left[\tilde{w}_l(\vec{k}) \cdot  \mathcal{F} \bigg\{\dfrac{I_l(\vec{r})}{\hat{S}_j (\vec{r}) \otimes h_l(\vec{r})} \otimes h_l(-\vec{r}) \bigg\} \right] \Bigg\}
\end{equation}
Note that the weights do not need to be re-computed throughout the iterations (as they are independent of $j$) and that the weighting can be incorporated directly into the second convolution operation:
\begin{equation}
	C_j(\vec{r}) =   \sum_{l=1}^2  \left[ \dfrac{I_l(\vec{r})}{\hat{S}_j (\vec{r}) \otimes h_l(\vec{r})} \otimes \left[w_l(\vec{r}) \otimes h_l(-\vec{r})\right] \right]
\end{equation}
with $w$ being the Fourier transform of the corresponding Fourier weight $\tilde{w}$. Note that this re-weighting of the "backpropagator" in the RL-algorithm is similar ot the work in \cite{Shroff_AcceleratingDeconvolution}. It differs in the sense that the weighting is only applied when multiple sub-images are available and the weighting $\tilde{w}_l$ is not dependent on any external parameters (it depends on the sub-PSFs only).\\
To test the weighted RL-algorithm the pupil splitting has been simulated on an artificial spokes target, with a splitting parameter $\eta = \sqrt{0.5}$. The individual sub-images (including Poissonian noise) are then recombined using multiview deconvolution (with and without weighting) and the corresponding full pupil (widefield) image by using the ordinary RL restoration algorithm. Note that we used the acceleration technique by Biggs et al. \cite{Biggs_RLAcceleration} and performed 20 iterations for all reconstructions. The achievable visibility $\mathcal{V}$ are shown in Fig. \ref{fig:FigA3a}:

\begin{figure}[htb]
	\centering
	\begin{subfigure}[c]{0.45\textwidth}
		\includegraphics[width=\linewidth]{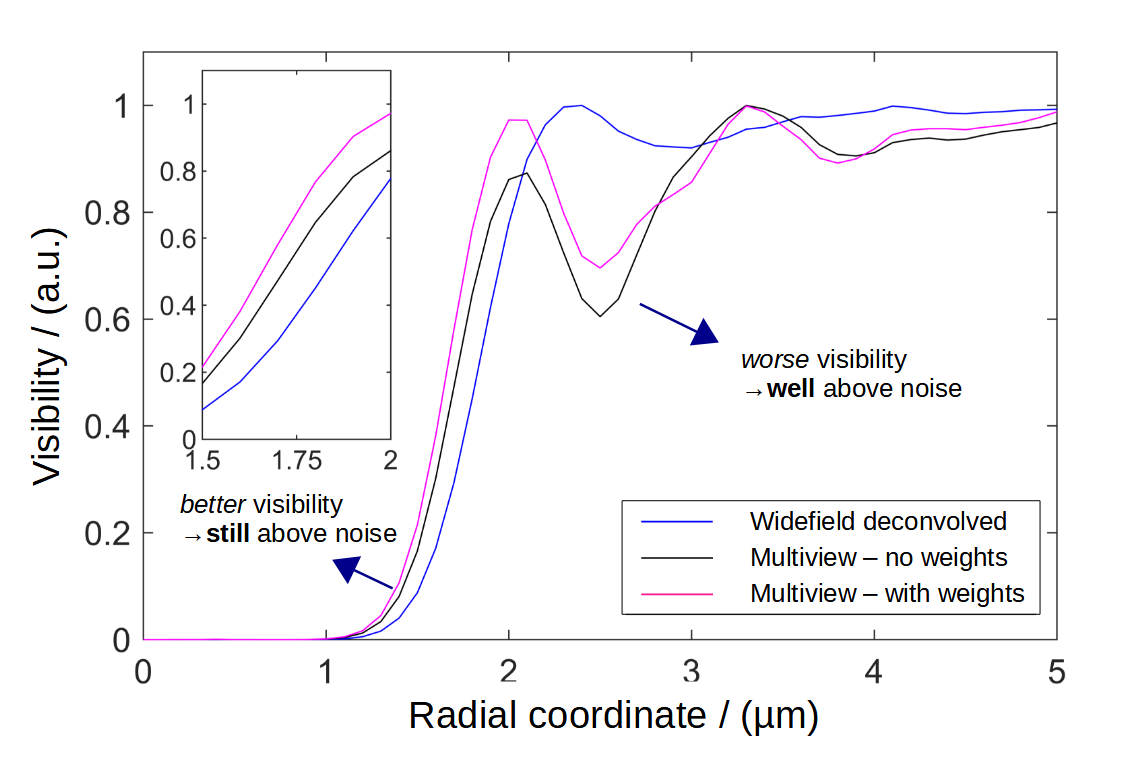}
		\caption{}
		\label{fig:FigA3a}
	\end{subfigure}
	\hspace{1cm}
	\begin{subfigure}[c]{0.45\textwidth}
		\includegraphics[width=\linewidth]{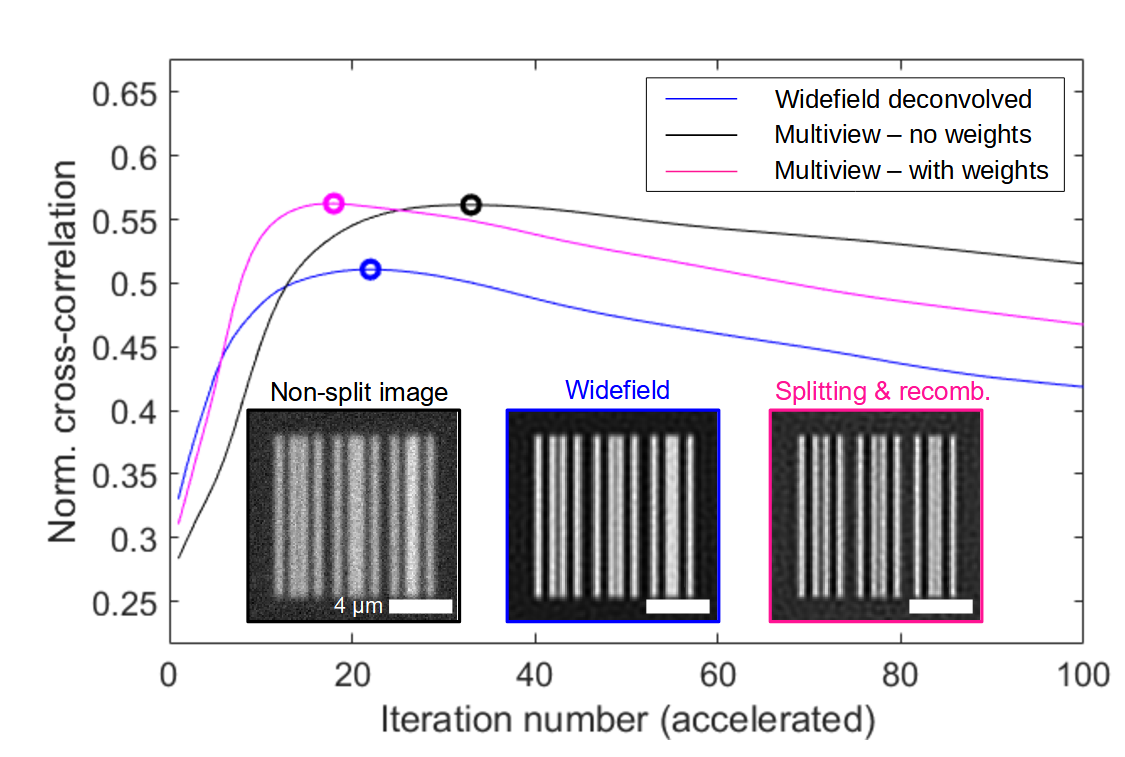}
		\caption{}
		\label{fig:FigA3b}
	\end{subfigure}
	\caption{\textcolor{blue}{(a)} Visibility value $\mathcal{V}$ in dependency of radius from the center of the simulated image of a spokes target.  While multiview deconvolution (black \& magenta) is not able to recover signal information at mid-frequencies with the same visibility as compared to the deconvolved WF result (blue), the achievable value of $\mathcal{V}$ is well above the noise (all reconstructions with $20$ iterations). Meaning that any object structure is still identifiable as such. Especially since high spatial frequencies are being recovered with larger $\mathcal{V}$ using the multiview approach. \textcolor{blue}{(b)} $\NCC$-curve (definition see text) of simulated line pairs (see inset) in dependency of the iteration number, for the single view (blue) and weighted multiview (black without weights; magenta with weights) deconvolution approach. Note that the optimum reconstruction result shows a larger $\NCC$-value for the multiview method, which is also reached earlier (18 iterations in case of multiview with weights, compared to 33 \& 24 for the multiview without weights \& single view reconstruction respectively). Even when the optimum iteration number is not known in an experimental setting, the $\NCC$-curve suggests that the benefit from using the multiview approach is maintained for all iterations.}
	\label{fig:FigA3}
\end{figure}

Note that the weighted multiview deconvolution (magenta) performs better at maintaining a larger $\mathcal{V}$ at the mid-frequencies, and even outperforms the WF deconvolution (blue) and non-weighted multiview reconstruction (black) at the highest spatial frequencies (be aware that we have not optimized the respective iteration number of each reconstruction). In terms of image visibility it is noteworthy that often only a minimum value of $\mathcal{V}$ is required to obtain a good estimate of the underlying object structure. Meaning that the dip at the mid-frequencies, in the pupil splitting approach, is not necessarily to be seen as a practical limitation.\\
Another point, which has not been properly taken care of in Fig. \ref{fig:FigA3a}, is the question whether both reconstructions, for non-split \& split data, has reached their individual optimum. Especially, introducing the re-weighting of the backpropagator does accelerate the convergence speed of the algorithm, while removing the convergence guarantee for which the RL-method is typically favored for. Note that this in principle is also true for the work in \cite{Shroff_AcceleratingDeconvolution}, but similar to them, we haven't noticed any stability issues with the weighted multiview deconvolution, yet. In terms of convergence speed we have analyzed the split pupil approach by comparing the simulated ground truth data with the deconvolved result for a varying number of iterations. This comparison is done by computing the normalized cross-correlation $\NCC$ \cite{Becker_PhD}:
\begin{equation}
	\NCC_j = \dfrac{\tfrac{1}{N} \sum_{\vec{r}} \left\{ \left[ S(\vec{r}) - \langle S (\vec{r}) \rangle \right] \cdot \left[ \hat{S}(\vec{r}) - \langle \hat{S}(\vec{r}) \rangle \right] \right\}}{\sigma_{S} \cdot \sigma_{\hat{S}}}
\end{equation}
with $\langle . \rangle$ being the average over all pixels, defined as (here for $S$):
\begin{equation}
	\langle S (\vec{r}) \rangle = \dfrac{1}{N} \sum_{\vec{r}} S(\vec{r}) 
\end{equation}
and $\sigma_{S,\hat{S}}$ being the standard deviation of $S$ and $\hat{S}$, calculated over $\vec{r}$. The $\NCC$ is a dimensionless number between $0 - 1$, which determines the accuracy of image restoration with respect to the ground truth $S$. It is plotted in case of simulated line pairs in Fig. \ref{fig:FigA3b} for the single view deconvolution of the WF-data (blue) and for the weighted multiview deconvolution of the split pupil approach (black without weights; magenta with weights). All three curves show a similar shape, but differ in the maximum achievable $\NCC$-value: the best reconstruction optimum is found after 18 iterations using the multiview algorithm with weighting, compared to 33 \& 24 iterations without the weights \$ using the WF-data alone. Also note that the reduction of reconstruction quality for further iterations goes with the same slope for both methods (single and weighted multiview deconvolution). Suggesting that even when the optimum iteration number is not chosen in an experimental setting, the multiview approach will still outperform the conventional single view deconvolution.

\subsection{Extended depth-of-field effect through pupil splitting}
\label{sec:EDoF_effect}

The axial extend $\Delta z$ of the WF-PSF $h_\WF$ is given according to \cite{Heintzmann_Sampling}:
\begin{equation}
	\Delta z = \dfrac{\lambda}{n \cdot [1 - \cos(\alpha)]}
\end{equation}
with $\alpha = \arcsin(\NA / n)$ the half-opening angle of the detection objective, $n$ the refractive index of the immersion medium and $\NA$ the numerical aperture. Hence we can express $\Delta z$ in terms of the numerical aperture according to:
\begin{equation}
	\Delta z = \dfrac{\lambda}{n}  \bigg[1 - \cos\bigg(\arcsin\left(\tfrac{\NA}{n}\right)\bigg)\bigg]^{-1}
\end{equation}
The trigonometric version of the Pythagorean theorem \cite{Bronstein_Maths} is given as:
\begin{equation}
	\sin^2(x) + \cos^2(x) = 1
\end{equation}
Which lets us express the cosine expression as:
\begin{eqnarray}
	\cos\bigg(\arcsin\left(\tfrac{\NA}{n}\right)\bigg) &=& \sqrt{1 -\bigg(\dfrac{\NA}{n}\bigg)^2}
\end{eqnarray}
Enabling us to write $\Delta z$ without any trigonometric functions, according to:
\begin{equation}
	\Delta z = \dfrac{\lambda}{n} \bigg[1 - \sqrt{1 - \left(\tfrac{\NA}{n}\right)^2}\bigg]^{-1}
\end{equation}
We can now calculate the axial extend of the PSF, corresponding to the  WF and disk pupil, as:
\begin{eqnarray}
	\Delta z_{\WF} &=& \dfrac{\lambda}{n} \bigg[1 - \sqrt{1 -\left(\tfrac{\NA}{n}\right)^2}\bigg]^{-1} \\
	\Delta z_{\Disk} &=& \dfrac{\lambda}{n} \bigg[1 - \sqrt{1 - \eta^2 \cdot \left(\tfrac{\NA}{n}\right)^2}\bigg]^{-1}
\end{eqnarray}
with the fact that $\mathcal{\tilde{P}}_\Disk$ corresponds to $\NA_\Disk = \eta \cdot \NA_\WF$. The axial extend of $h_\Ring$ can be indirectly computed from $\Delta z_\WF$ and $\Delta z_\Disk$ by making use of $\mathcal{\tilde{P}}_\Ring = \mathcal{\tilde{P}}_\WF - \mathcal{\tilde{P}}_\Disk$. This enables us to estimate the axial extend of $\mathcal{\tilde{P}}_\Ring$ in Fourier space as $\Delta k_{z,\Ring} = \Delta k_{z,\WF} - \Delta k_{z,\Disk}$. Going back into real space, which relates $\Delta k_{z,\Ring}$ to $\Delta z_\Ring$, gives us:
\begin{equation}
	\Delta z_\Ring = \dfrac{\Delta z_\WF \cdot \Delta z_\Disk}{\Delta z_\Disk - \Delta z_\WF}
\end{equation}
For the experimental case ($\NA_\WF = 1.2$, $n = 1.333$, $\lambda \approx 520$ nm, $\eta = \sqrt{0.5}$) we obtain the following values for the axial extend of the individual PSFs:
\begin{eqnarray}
	\Delta z_\WF &=& 664 \text{   nm} \\
	\Delta z_\Disk &=& 1640 \text{   nm} \\
	\Delta z_\Ring &=& 1117 \text{   nm}
\end{eqnarray}
Indicating that splitting the pupil of a widefield detection system into two sub-pupils with equal area ($\eta = 0.5$), will enlarge the depth-of-field the new approach by approximately a factor of two.

\subsection{Results of imaging the resolution target}
\label{sec:Image_results_ResolutionTarget}

\begin{figure}[htb]
	\centering
	\includegraphics[width=0.8\linewidth]{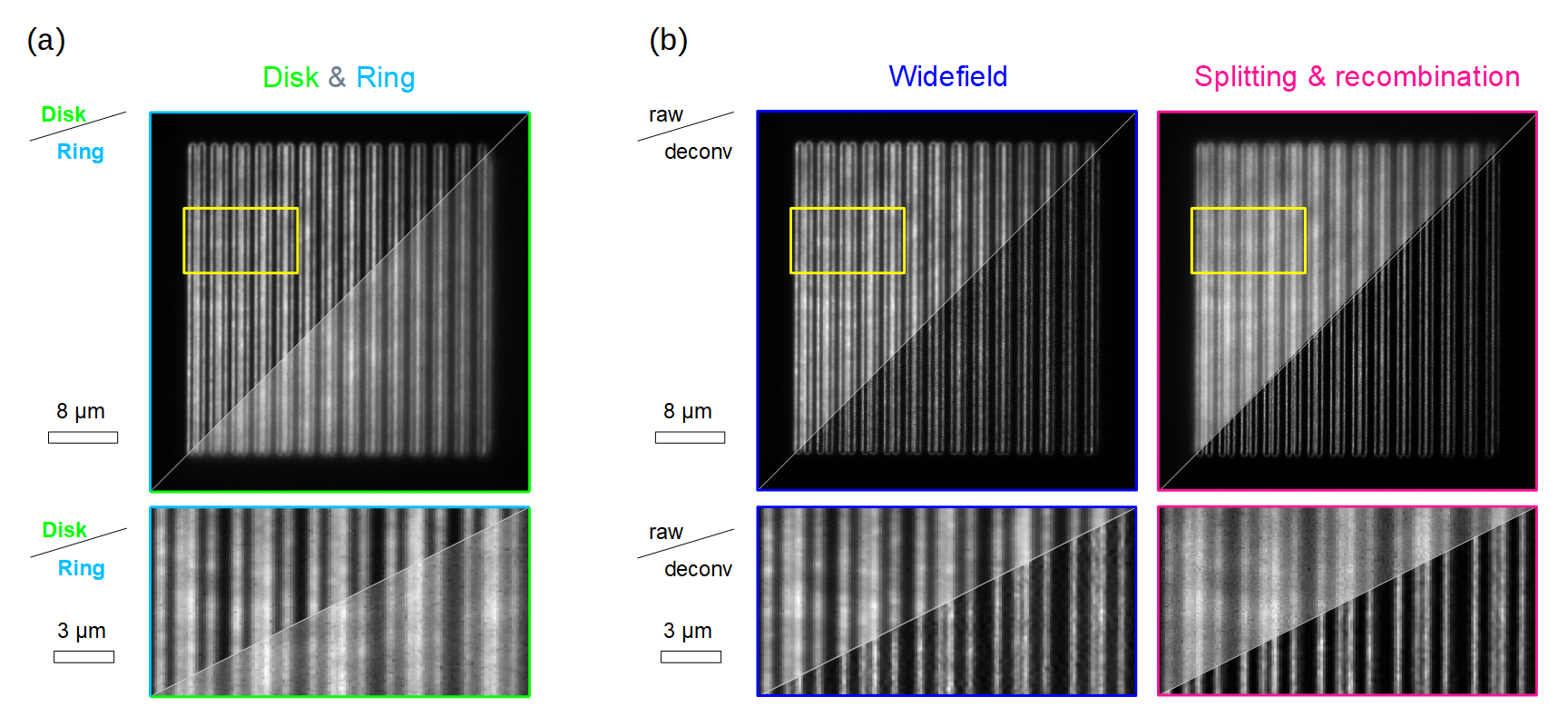}
	\caption{Results of imaging the resolution target of the Argolight calibration sample \cite{Royon_Argolight}. \textcolor{blue}{(a)} Split image data corresponding to a splitting parameter of $\eta = 0.68$. Note how the image corresponding to $\mathcal{\tilde{P}}_{\Ring}$ visually seems to exhibit stronger blur, nevertheless maintains a detectable modulation for the line patterns very close to the cut-off frequency (in zoom) of the detection system. \textcolor{blue}{(b)} Widefield and split pupil data, before and after deconvolution (WF = 24; SP = 18 iterations). Visually both results look similar, a look at the corresponding line profiles (weighted averaging in Fig. \ref{fig:FigA5a}; multiview deconvolution in Fig. \ref{fig:Fig4}) however shows an improvement for the pupil split approach (also see Tab. \ref{tab:Tab2}).}
	\label{fig:FigA4}
\end{figure}

The experimental results of imaging the Argolight resolution target with and without splitting are shown in Fig. \ref{fig:FigA4}. The split data is depicted in \textcolor{blue}{(a)}, where the zoom (bottom row) shows a region close to the resolution limit of the detection system. On a first glance the image corresponding to $\mathcal{\tilde{P}}_{\Ring}$ (bottom right) looks more blurry, nevertheless, the very fine structures (the line pairs of interest)  are actually better resolvable than the $\mathcal{\tilde{P}}_{\Disk}$ case (top left). This demonstrates that this high spatial frequency information can actually be transferred with less photons than the WF-case (left in \textcolor{blue}{(b)}), using the ring pupil.

The widefield and split \& recombined image results are shown before and after deconvolution (WF = 24; SP = 18 iterations) in Fig. \ref{fig:FigA4} \textcolor{blue}{(b)}. Visually both results (WF and SP) look similar, but comparing the two line profiles in Fig. \ref{fig:Fig4} and Fig. \ref{fig:FigA5a} shows an improvement for the split pupil case. In case of weighted averaging it can be seen that the curve corresponding to splitting \& recombination approach (magenta) is reduced in its peak values. This is due to the missed interference between light from $\mathcal{\tilde{P}}_{\Disk}$ and $\mathcal{\tilde{P}}_{\Ring}$. Nevertheless, the modulation corresponding to the very closely space line pairs is improved. Even such, that some of the patterns which previously (with WF detection) were not resolved, could now be identified as a line pair (indicated with $\infty$ in Tab. \ref{tab:Tab2})).

\if
\subsection{The aplanatic factor modifying the pupil splitting}
\label{sec:Aplanatic_factor}

An effect which has been neglected in the previous theoretical analysis of pupil splitting is that of the \emph{aplanatic factor} \cite{Goodman_Fourier}. Which simply states that light emitted by a point source does not yield a pupil with a constant radial energy density, but shows a slight increase towards the rim of the BFP. This effect is purely of geometrical nature as an emitted, infinitesimally small, light cone has to be projected at the \emph{Gaussian reference sphere} \cite{Mertz_Book} to be matched to $\mathcal{\tilde{P}}_\WF$. The projection gets stronger the larger the angle of emission $\alpha$ and is given as $\gamma$:
\begin{equation}
	\gamma(\alpha) = \dfrac{1}{\cos \left(\alpha\right)}
\end{equation}
The radial position in the BFP is given as $r = \NA_\WF \cdot f_{obj.}$, with $f_{obj.}$ being the focal length of the detection objective.\\
We can use this to re-define the aplanatic factor $\gamma$ as:
\begin{equation}
	\gamma(r) = \dfrac{1}{\cos \left[\arcsin \left(\tfrac{r}{n \cdot f_{obj.}}\right)\right]}
\end{equation}
Because the BFP of the detection system will carry more and more photons towards the edge, a splitting with $\eta = \sqrt{0.5}$ will not lead to two sub-images carrying the same amount of photons. We can show this by computing the radial energy density $E(r_\Min,r_\Max)$:
\begin{equation}
	E(r_\Min,r_\Max) = \int_{0}^{2\pi} d\varPhi \int_{r_\Min}^{r_\Max} dr \hspace{4pt} r \cdot \gamma(r) = 2\pi \cdot \left(f_{obj.}n\right)^2 \cdot \left[1 - \sqrt{1 - \left(\tfrac{r}{f_{obj.}n}\right)^2}\right]_{r_\Max}^{r_\Min}
\end{equation}
With this, we are now interested in the ratio $E(0,R_\Disk) / E(R_\Disk,R_\WF)$, which we have calculated in the case of $R_\Disk = \eta \cdot R_\WF$ as:
\begin{equation}
	\dfrac{E(0,R_\Disk)}{E(R_\Disk,R_\WF)} = \dfrac{0.2740}{0.7260} = 0.3774
\end{equation}
with $f_{obj.} = 3$ mm, $n = 1.333$, $R_\WF = 4.0$ mm, $\eta = 0.68$ and $R_\Disk = 2.75$ mm. What this does is to effectively reduce the splitting parameter $\eta$ to a new value of $\eta = \sqrt{0.3774} \approx 0.61$. Meaning that the experimentally realized pupil split does not yield an equal area split, and  therefore changes the region of improvement $\IF > 0$ by setting $k_0 = 0.5 \cdot (1 + 0.61) /k_\WF = 0.805/k_\WF$. Which still is not enough to fully explain why we see an improvement of $\mathcal{V}$ already at $|\vec{k}|/k_0 = 0.66$. Meaning that probably the optical system is not diffraction-limited.
\fi

\subsection{Results of imaging the stairs target}
\label{sec:Image_results_StairsTarget}

\begin{figure}[htb]
	\centering
	\begin{subfigure}[c]{0.45\textwidth}
		\includegraphics[width=\linewidth]{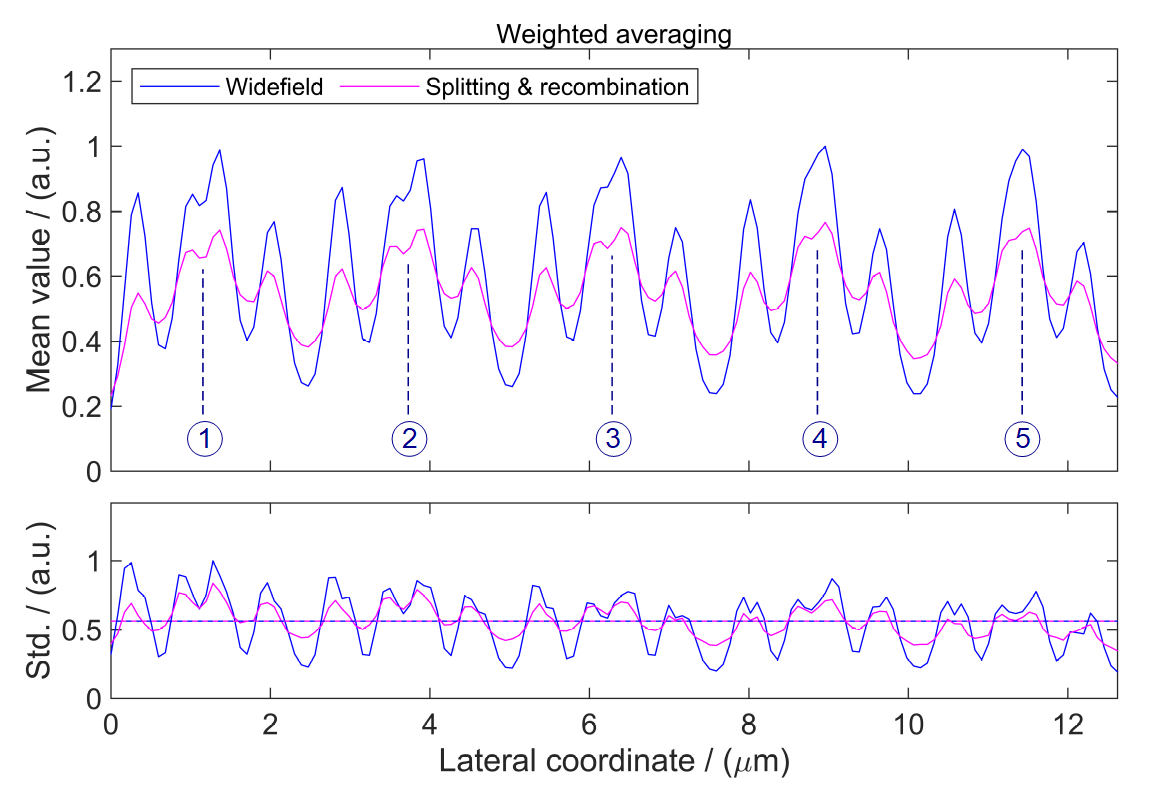}
		\caption{}
		\label{fig:FigA5a}
	\end{subfigure}
	\hspace{1cm}
	\begin{subfigure}[c]{0.45\textwidth}
		\includegraphics[width=\linewidth]{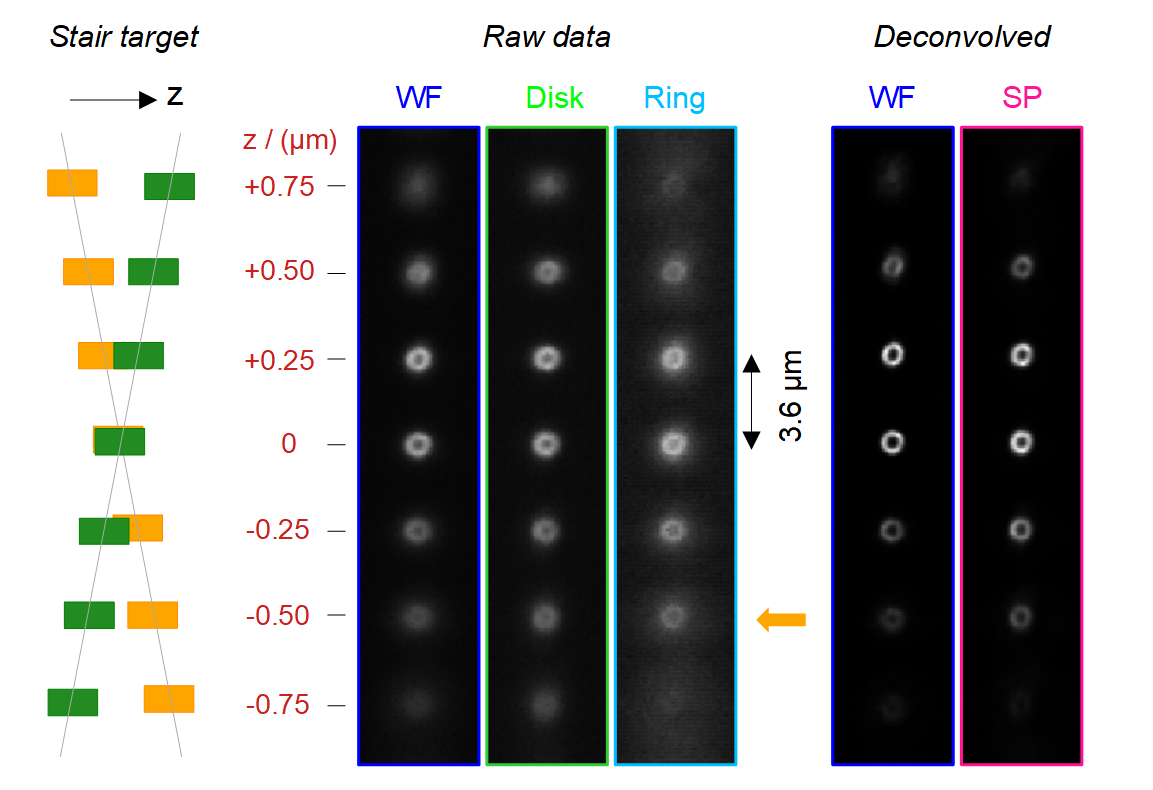}
		\caption{}
		\label{fig:FigA5b}
	\end{subfigure}
	\caption{\textcolor{blue}{(a)} Line profiles for imaging the resolution target using a full pupil (WF, blue) and by splitting the pupil ($\eta = 0.68$) and performing weighted averaging recombination (magenta). For the very fine line pairs the WF result shows no modulation, while the split pupil result still maintains some visibility $\mathcal{V}$ (also see Tab. \ref{tab:Tab2}). \textcolor{blue}{(b)} Imaging the stair target illustrates the extended depth-of-field achieved through the split pupil approach. A schematic of the sample is shown on the left, indicating the two crossing stairs (green and orange) with a stair step size of $0.25$ $\mu$m. The raw image data is shown in the middle, where the orange arrow indicates a stair step which is still resolvable after splitting, but not using the full pupil. After deconvolution (right) we can see that the aforementioned stair is reconstructed more strongly than in the WF case (here 18 iterations for both). It seems that multiview deconvolution (magenta) is able to reconstruct more information although only a 2D deconvolution was applied.}
	\label{fig:FigA5}
\end{figure}

The actual ability to generate an extended depth-of-field was not the main goal behind our work. Nevertheless, it can be experimentally studied using the stair sample (pattern I) from the Argolight calibration target \cite{Royon_Argolight}. Figure \ref{fig:FigA5b} left shows a schematic of the two crossing stairs (in green and orange), indicating the different stair steps with a step size of $0.5$ $\mu$m. 

The focal plane of the microscope system in Fig. \ref{fig:Fig5} was adjusted such that the center part of the stair pattern was in-focus. The recorded WF data (blue) in Fig. \ref{fig:FigA5b} middle shows that the stair steps gradually defocuses. As expected, the images corresponding to $\mathcal{\tilde{P}}_\Disk$ and $\mathcal{\tilde{P}}_\Ring$ exhibit a larger depth-of-focus, meaning that more out-of-focus steps can still be resolved (e.g. see orange arrow).\\
Additionally we used the multiview deconvolution approach to recombine the (2D) data, the results are shown on the right of Fig. \ref{fig:FigA5b} (here 18 iterations for both methods). Again, the out-of-focus "steps" are more strongly present in the multiview reconstruction result. However, it seems that resolving the ring shape is still slightly better with the WF approach. With split-pupil imaging a potentially powerful approach would be to estimate 3D information  from the two split-pupil images by an approach similar to \cite{Jost_ThickSlice,Becker_PhD}. In this way a 2D measurement yields three-dimensional information through deconvolving it with a 3D point-spread function, which either has to be measured or modeled. For more detail see \cite{Becker_PhD}.


\end{document}